\RequirePackage{ifpdf}
\documentclass[hyper]{JHEP3}

\pdfoutput=1
\usepackage{epsfig}
\usepackage{cite}
\usepackage{graphicx}
\usepackage{subfigure}
\usepackage{inputenc}
\usepackage{amsmath}
\usepackage{slashed}
\usepackage[ruled,norelsize]{algorithm2e}
\usepackage{empheq}
\usepackage{cancel}
\usepackage{multirow}

\newcommand{\beq}{\begin{equation}}
\newcommand{\eeq}{\end{equation}}
\newcommand{\beqn}{\begin{eqnarray}}
\newcommand{\eeqn}{\end{eqnarray}}
\def\df{{\rm d}}
\newcommand{\bn}{{\bar n}}

\newcommand{\mcdot}{\!\cdot\!}
\newcommand{\eq}[1]{Eq.~\eqref{eq:#1}}
\newcommand{\eqs}[2]{Eqs.~\eqref{eq:#1} and \eqref{eq:#2}}

\newcommand{\fullF}{\mathcal{F}}

\def\eps{\delta}

\def\nn{\nonumber\\}

\newcommand{\w}{\omega}
\newcommand{\W}{\Omega}
\newcommand{\vk}{\vec{k}}
\newcommand{\Df}{\mathcal{D}}
\newcommand{\abs}[1]{\lvert#1\rvert}

\newcommand{\nbar}{{\bar{n}}}
\newcommand{\tdot}{\!\cdot\!}

\def\psl{p\!\!\!\slash}
\def\ksl{k\!\!\!\slash}

\newcommand{\ket}[1]{\lvert#1\rangle}
\newcommand{\cE}{ \mathcal{E} }
\newcommand{\cL}{ \mathcal{L} }
\newcommand{\vn}{\vec{u}}

\newcommand{\vtS}{\tilde V_s}
\newcommand{\vtn}{\tilde V_n}

\newlength\dlf  

\title{A formalism for the resummation of non-factorizable observables in SCET}

\author{Christian W.~Bauer$^{a}$ and Pier Francesco Monni$^b$\\
   $^a$Ernest Orlando Lawrence Berkeley National Laboratory, University of California, Berkeley, CA 94720, USA\\
   $^b$CERN, Theoretical Physics Department, CH-1211 Geneva 23, Switzerland\\   
      E-mail: \email {cwbauer@lbl.gov}, \email{pier.monni@cern.ch}
        }

\received{\today}               
\accepted{\today}               

\preprint{CERN-TH-2019-098}

\abstract{In the framework of soft-collinear effective theory (SCET),
  we show how to formulate the resummation for a broad family of
  final-state, global observables in high-energy collisions in a
  general way that is suitable for a numerical calculation. Contrary
  to the standard SCET approach, this results in a method that does
  not require an observable-specific factorization theorem. We present
  a complete formulation at next-to-next-to-leading logarithmic order
  for $e^+e^-$ observables, and show how to systematically extend the
  framework to higher orders. This work paves the way to automated
  resummation in SCET for several physical observables within a single
  framework.}


\begin{document} 

\newpage
\section{Introduction}
\label{sec:introduction}
Perturbation theory is one of the most widely used techniques to make
predictions for interacting quantum field theories. In fixed order
(FO) perturbation theory, one expands the physical observables in
powers of the coupling constants of the theory, where leading order
(LO) predictions describe a given process to lowest order, while
higher order corrections are suppressed by additional powers of
$\alpha_i = g_i^2 / 4\pi$. 
The dominant corrections are in most cases due to the strong
interaction, where the expansion is in terms of $\alpha_s$. This
approach is well suited for sufficiently inclusive observables. On the
other hand, for observables that depend on several scales, it is well
known that the perturbative expressions contain logarithmic dependence
on the ratio of these scales $r$, with up to two powers of such
logarithms for each power of the coupling constant
($\alpha_s^n \ln^{2n} r$). In this case the FO perturbative
expressions can become unreliable and a different expansion is
required to make accurate predictions. Resummed perturbation theory is
a rearrangement of the perturbative series such that the dominant
logarithmic terms are resummed to all orders in the coupling constant
$\alpha_s$.

In this article we consider the family of global, recursively infrared
and collinear (rIRC) safe~\cite{Banfi:2004yd} observables $v$ whose
leading order perturbative series features double logarithms in the
limit $v\to 0$. For such observables, the perturbative series of the
cumulative distribution
\begin{align}
\Sigma(v) &= \frac{1}{\sigma_B}\int_0^v\df v' \frac{d\sigma}{d v'}\,,
\end{align}
with $\sigma_B$ being the Born cross section, can be shown to
exponentiate at the leading logarithmic order such that one can write
\begin{align}
\label{eq:SigmaGeneralForm}
\Sigma(v) = \exp\left[ L g_0(\alpha_s L) + g_1(\alpha_s L) + \alpha_s g_2(\alpha_s L) + \ldots\right]
\,,
\end{align}
where
\begin{align}
L = \ln \frac{1}{v}
\,.
\end{align}
In the limit $L \sim 1 / \alpha_s$, such that $\alpha_s L \sim 1$,
each term in a Taylor expansion of the functions $g_i(\alpha_s L)$ is
of the same size, such that arbitrarily large powers of $\alpha_s$ are
required to determine these functions. However, the contribution of
the function $g_n$ is suppressed by one power of $\alpha_s$ compared
to the function $g_{n-1}$.  The goal of resummation is to compute
these functions, and the N$^k$LL order is defined by the inclusion of
the functions $g_{m}(\alpha_s L)$ with $m \le k$ in the prediction
(e.g. at NLL one needs $g_0(\alpha_s L)$ and $g_1(\alpha_s L)$).

Resummed expressions at NNLL and beyond have been derived for a wide
range of observables both for
$e^+e^-$~\cite{Becher:2008cf,Chien:2010kc,Becher:2012qc,Hoang:2014wka,Becher:2015lmy,Banfi:2014sua,Frye:2016aiz,Banfi:2016zlc,Moult:2018jzp,Procura:2018zpn,Banfi:2018mcq}
as well as in hadronic
collisions~\cite{Bauer:2002nz,Banfi:2004nk,Bozzi:2005wk,Becher:2010tm,Stewart:2010pd,
  Banfi:2011dx,Berger:2010xi,Jouttenus:2011wh
  ,Becher:2012qa,Zhu:2012ts,Banfi:2012du,Banfi:2012jm,Becher:2013xia,Stewart:2013faa,Procura:2014cba,Banfi:2015pju,Monni:2016ktx,Bizon:2017rah,Chen:2018pzu,Bizon:2018foh,Catani:2018mei,Lustermans:2019plv}.

Two main approaches to calculate the resummed expressions exist. The
branching
formalism\cite{Catani:1990rr,Catani:1991kz,Banfi:2004yd,Banfi:2014sua,Banfi:2018mcq}
uses the factorization properties of squared amplitudes in QCD to
describe the radiation dynamics at all perturbative orders. This
approach is widely used also in parton shower algorithms, which are
typically accurate to (N)LL for specific
observables~\cite{Catani:1990rr,Nagy:2009vg,Dasgupta:2018nvj,Bewick:2019rbu}. However,
in contrast to a parton shower, in a resummed calculation one does not
need to simulate physically sensible events which exhibit momentum
conservation. This makes the branching formalism usable in principle
to any logarithmic accuracy. 
The main idea relies on the computation of the squared amplitudes with
an arbitrary number of soft or collinear emissions, but requiring them
to be correct only to a given logarithmic accuracy. With these
simplifications one can formulate a Monte-Carlo (MC) algorithm to
simulate the radiation above a certain resolution scale $q_0$, while
the emissions below $q_0$ are treated analytically. For sufficiently
simple observables one can rewrite the branching formalism in terms of
differential equations, which can be solved in closed form.

An alternative approach is based on the derivation of a factorization
theorem for the observable under consideration, in which the cross
section is decomposed in terms of ingredients that are sensitive to a
given infrared scale. Logarithms are then resummed through evolution
equations of the various ingredients. Collins, Soper and
Sterman~\cite{Collins:1984kg} pioneered this technique with their
seminal calculation of the transverse momentum distribution of the
vector boson in $W$ and $Z$ production. Subsequently, this approach
was formulated~\cite{Bauer:2002nz} in soft-collinear effective theory
(SCET)~\cite{Bauer:2000ew,Bauer:2000yr,Bauer:2001ct,Bauer:2001yt}. In SCET, the interactions between particles can be factorized at the
level of the Lagrangian~\cite{Bauer:2001ct} and, in the case of
observables with factorizing measurement
functions~\cite{Bauer:2008dt,Bauer:2008jx}, one can separate the cross
section into contributions from hard, collinear and soft degrees of
freedom.
In dimensional regularization, each element of the factorization
theorem then only depends on a single scale through its ratio to the
renormalization scale~\cite{Bauer:2000ew,Bauer:2000yr}. Therefore, the
renormalization (RG) group evolution can be used to resum the
logarithmic dependence of each of the ingredients in the factorization
theorem, and the resummation is achieved via the computation of
anomalous dimensions of SCET operators.

Using the thrust observable as a case study, in~\cite{Bauer:2018svx}
it was shown how these two formulations of resummation can be combined
into a hybrid method. This approach proceeds by defining a simpler
version of thrust, defined in order to obey a particularly simple
factorization theorem that can be handled using the standard SCET
formalism. In the resulting factorisation theorem, the different
ingredients are combined through a multiplication, rather than a more
complicated convolution as in the case of the full thrust observable,
hence leading to a much simpler structure of the RG evolution. 

Given a resummed result for the the cumulative distribution of the
simplified thrust observable $\Sigma_{\rm max}(\tau)$, one could then
obtain the actual thrust cumulative distribution $\Sigma(\tau)$ as
\begin{align}
\label{eq:formula_thrust}
\Sigma(\tau) = \Sigma_{\rm max}(\tau) \,\fullF(\tau)
\,,
\end{align}
where the transfer function $\fullF(\tau)$ can be calculated
numerically using an efficient MC algorithm.

In~\cite{Bauer:2018svx} it was argued that this procedure is rather
general, and can be applied to a wide variety of observables, without
giving the details of this generalization. The purpose of this paper
is to derive this formulation for a generic recursive infrared and
collinear (rIRC) safe observable $v$ and to give all the details
required for a numerical implementation up to NNLL accuracy. In
particular, we derive the generalization of \eq{formula_thrust} namely
\begin{align}
  \Sigma(v) = \Sigma_{\rm max}(v) \,\fullF(v)\,
  \,,\end{align}
give a particularly general definition of the simplified observable $\Sigma_{\rm max}(v)$, and provide a detailed derivation of the transfer function $\fullF(v)$.

The paper is organized as follows: We start in
Section~\ref{sec:logcounting} with some comments about the counting of
the logarithms in a resummed expressions beyond NLL accuracy. After
defining the kinematics of the process and detailing the notation used
throughout the paper in Section~\ref{sec:notation}, we introduce the
basic idea behind the numerical resummation in
Section~\ref{sec:basic_idea}. In that section, we briefly touch on all
aspects of the derivation: we introduce a generic and fully
differential factorization approach derived from SCET, from which any
observable can be calculated, define the simple observable and how it
can be resummed using SCET and finish by an SCET definition of the
transfer function. In Section~\ref{sec:transfer_fully_differential} we
then derive a general expression for a fully differential transfer
function, from which the transfer function for any observable can be
calculated. While in Section~\ref{sec:transfer_fully_differential} the
derivation is purposely kept rather general, we give explicit and
detailed results for the transfer function in
Section~\ref{sec:explicit_transfer} at both NLL and NNLL accuracy. In
Section~\ref{sec:observable} we discuss various ways one can simplify
the observable dependence in the transfer function. We present our
Conclusions in Section~\ref{sec:Conclusions}.

\section{Remarks on the counting of logarithms}
\label{sec:logcounting}

In \eq{SigmaGeneralForm} we presented the general form of the resummed
expression for a cumulative distribution of a general observable. To
obtain N$^k$LL accuracy one needs to include the functions
$g_n(\alpha_s L)$ with $n \leq k$, which ensure a reliable theory
description in the regime $L \sim 1 / \alpha_s$. Within this counting,
NLL accuracy is defined by including the functions $g_0(\alpha_s L)$
and $g_1(\alpha_s L)$, which are required to obtain a prediction that
is accurate up to ${\cal O}(\alpha_s)$ corrections. Beyond NLL
accuracy, one includes the functions $g_n(\alpha_s L)$ with
$n \geq 2$, which lead to progressively suppressed perturbative
corrections.

This allows for different ways of consistently defining the resummed
cross section beyond NLL, such that they differ from one another by
terms beyond the considered order. Consider as an example the case of
a NNLL resummation. The first scheme is to keep the full exponential
form of \eq{SigmaGeneralForm}, which we repeat here for convenience:
\begin{align}
\label{eq:scheme1}
\Sigma_{\rm NNLL}^{(1)} = \exp\left[ L g_0(\alpha_s L) + g_1(\alpha_s L) + \alpha_s g_2(\alpha_s L)  \right]
\,.
\end{align}
A second counting scheme can be defined by expanding the perturbative corrections introduced by the $g_2$ term in \eq{scheme1} in
powers of $\alpha_s$, but keeping the scaling $L \sim 1 / \alpha_s$
intact. This allows one to write
\begin{align}
\label{eq:scheme2}
\Sigma_{\rm NNLL}^{(2)} = \exp[L g_0(\alpha_s L) + g_1(\alpha_s L)] \left[ 1 + \alpha_s g_2 (\alpha_s L) + \ldots \right]\,.
\end{align}
It is easy to verify that the difference between
eq.~\eqref{eq:scheme2} and eq.~\eqref{eq:scheme1} amounts to terms
\begin{align}
\Delta \Sigma_{\rm NNLL} = {\cal O}(\alpha_s^2) \times \exp[L g_0(\alpha_s L) + g_1(\alpha_s L)]
\,,
\end{align}
which contribute only to N$^3$LL.

Finally, one can define a third scheme, which is a hybrid between the
two schemes defined so far. In this scheme one writes
\begin{align}
\label{eq:scheme3}
\Sigma_{\rm NNLL}^{(3)} = \exp[L g_0(\alpha_s L) + g_1(\alpha_s L) + \alpha_s \tilde g_2(\alpha_s L)] \left\{ 1 + \alpha_s \left[g_2 (\alpha_s L) - \tilde g_2(\alpha_s L) \right] + \ldots \right\}
\,,
\end{align}
where $\tilde g_2(x) \neq g_2(x)$. Therefore, this scheme keeps part
of the NNLL corrections in the exponent, and expands out the
remainder. Once again the difference between eq.~\eqref{eq:scheme3}
and the previous two schemes is of order N$^3$LL.

All of these counting schemes give equivalent predictions if
$\alpha_s L \sim 1$, since they only differ by terms that are of
higher order than those considered.
Generally, most observables do not naturally satisfy the fully
exponentiated form of eq.~\eqref{eq:scheme1}, and some NNLL
contributions (and beyond) are usually expanded out in a fashion
similar to scheme~\eqref{eq:scheme3}. However, the exact definition of
the function $\tilde g_2(\alpha_s L)$ is intrinsically ambiguous, in
that one can always reshuffle NNLL terms from the exponent to the
expanded remainder and vice versa.
More precisely, different approaches to resummation lead to different
definitions of the exponentiated $\tilde g_2(\alpha_s L)$ function,
while yielding the same NNLL result for the resummed cross
section. For instance, standard SCET resummations keep all constant
terms and their own running-coupling corrections outside of the
exponent, while some keep part of the constants (specifically those
coming from the virtual corrections) exponentiated. In the approach of
refs.~\cite{Banfi:2014sua,Banfi:2016zlc,Banfi:2018mcq}, conversely,
only the universal contributions $\tilde g_2(\alpha_s L)$ are
exponentiated, while the observable-dependent NNLL corrections are
expanded out at fixed $\alpha_s L$.

This freedom can be exploited in the formulation of numerical
approaches to resummation. This can easily be understood from the way
numerical resummation approaches work by relating two different
observables that have the same LL structure. Given two such
observables $\Sigma$ and $ \Sigma^\prime$, whose logarithmic
resummation is defined by the functions $g_i$ and $g^\prime_i$ with
$g_0 = g^\prime_0$, the two resummations can be related as
\begin{align}
\Sigma^{ \rm NNLL}(v) = &\Sigma^{\prime \rm
                        NNLL}(v)\,\frac{\Sigma^{\rm
                        NNLL}(v)}{\Sigma^{\prime \rm NNLL}(v)} =
                        \exp\left[L g^\prime_0(\alpha_s L) +
                        g^\prime_1(\alpha_s L) +\alpha_s g^\prime_2(\alpha_s L)\right] \notag\\
& \times \exp\left[g_1(\alpha_s L) - g^\prime_1(\alpha_s L) \right] \left[1 + \alpha_s \left(g_2(\alpha_s L) -
  g^\prime_2(\alpha_s L)\right) \right]+ {\cal O}({\rm N^3LL})
\,.
\end{align}
This means that the difference between the $g_2$ and $g_2^\prime$
functions, rather than being exponentiated, can be evaluated at fixed
order while keeping $\alpha_s L$ fixed. This will considerably
simplify the numerical algorithm. This feature has been exploited to
formulate approaches that push the automatic NLL resummation of {\tt
  CAESAR}~\cite{Banfi:2004yd} beyond NLL accuracy, in particular in
the algorithms adopted in the programs {\tt ARES}~\cite{Banfi:2014sua}
or {\tt RadISH}~\cite{Monni:2016ktx,Bizon:2017rah}. The above
conclusions will be used in this paper to formulate a resummation
method in the framework of SCET that can be implemented in a numerical
algorithm.  Note that the same considerations apply to any logarithmic
accuracy beyond NLL, for which different resummation approaches
predict resummed cross sections that differ by corrections beyond the
considered order. It follows that a comparison between resummed
predictions must be performed up to the nominal logarithmic order,
while subleading differences may well be present.

\section{Kinematics and notation}
\label{sec:notation}

In this section we introduce a convenient notation to formulate the
resummation in a way that is applicable to broad classes of
observables. We consider $e^+e^-$ collisions producing a $Z/\gamma^*$
which subsequently decays into hadrons, i.e.
$e^+e^- \to q_n q_\bn + X$. We begin by introducing the usual
light-cone parametrization of four-momenta with respect to a
light-like direction $n^\mu = (1, \vec{n})$ with $n^2 = 0$. This
allows one to write
\begin{align}
k^\mu = \frac{n^\mu}{2} \bn \cdot k + \frac{\bn^\mu}{2} n \cdot k + k_t^\mu 
\equiv \frac{n^\mu}{2} k^- + \frac{\bn^\mu}{2} k^+ + k_t^\mu 
\,,
\end{align}
where $\bn^\mu = (1, -\vec{n})$, such that $\bn^2 = 0$,
$n \cdot \bn = 2$.  The momentum $k_t$ is a space-like momentum
satisfying $n \cdot k_t = \bn \cdot k_t = 0$. We also define
$k_\perp^2 = - k_t^2$. A common choice is to align $\vec{n}$ with
the $z$-axis, resulting in $n = (1, 0, 0, 1)$, but the notation is
generic.

Next, we introduce the notation that allows us to describe the fully
differential energy distribution of the considered process. We
consider a state $X$ containing $M$ massless particles with four-momenta
$k_i = (E_i, \vec{k}_i)$. The total energy $\omega_X$ and 3-momentum
$\vec{k}_X$ at a given solid angle $\Omega$ in the final state are
given by
\begin{equation} \label{eq:wXkX}
\w_X(\W) = \sum_{i = 1}^M E_i\, \delta(\W - \W_i)
\,,\qquad
\vk_X(\W) = \sum_{i = 1}^M \vec{k}_i\, \delta(\W - \W_i)
\,,\end{equation}
where the solid angle of each particle is determined by its 3-momentum $\W_i = \W(\vec{k}_i)$. 
One can now define a functional integration measure by discretization. We divide $\W$ into infinitesimally small bins $\{\W_k\}$, and define the set of discrete variables $\{\w_k,\vk_k\}$ as the integrals of $\w(\W)$ and $\vk(\W)$ over the bins $\{\W_k\}$,%
\begin{equation}
\w_{\W_k} = \int_{\W_k}\!\df\W\, \w(\W)
\,,\qquad
\vk_{\W_k} = \int_{\W_k}\!\df\W\, \vk(\W)
\,.\end{equation}
Now we define an integration measure over the energies and 3-momenta at a given solid angle $\W_k$ as
\begin{align}
[\df \w]_{\W_k} = \df \w(\W_k) \, \Theta(\w(\W_k))\,, \qquad [\df^3 k]_{\W_k} = \df^3 k(\W_k) \, \delta^2( \W(\vk(\W_k))- \W_k)
\,.
\end{align}
Restricting the particles inside of each solid angle to be on-shell (which is justified since we have infinitesimally small solid angles $\W_k$) one can write
\begin{align}
[\df k]_{\W_k} = \frac{[\df \w]_{\W_k} \, [\df^3 k]_{\W_k}}{(2\pi)^3} \,  \delta( \w(\W_k)^2- \vec{k}^2(\W_k)) = \frac{\w_{\W_k} \, [\df \w]_{\W_k}}{2(2\pi)^3}
\,.
\end{align}

With this notation, the phase space for a state $X_N$, containing $N$ final state particle plus the two Born particles $q_n$ and $q_\bn$ is given by 
\begin{align}
\label{eq:Dwn}
\Df \w^{(N)} = S(N) [\df q_n]\,[\df q_\bn] \prod_{i=1}^N \df \W_i \, [\df k]_{\W_i} \equiv S(N)  [\df q_n]\,[\df q_\bn]\prod_{i=1}^N [\df k_i]
\,,
\end{align}
where the symmetry factor $S(N)$ takes into account identical particles in the final state (for example, for $N$ gluons we have $S(N) = 1 / N!$). 

Since a given energy density is identified by its momenta, as discussed above, the sum of two energy distributions is simply given by the the combined momenta of the two individual distributions
\begin{align}
\label{eq:omega_union}
\w = \w_1 + \w_2 = \left \{ k_{i,1} \right\} \cup \left \{ k_{i,2} \right\} 
\,.
\end{align}
One simple consequence of this result is that the total momentum is conserved
\begin{align}
P^\mu[\w] = P^\mu[\w_1] + P^\mu[\w_2]
\,,
\end{align}
where $P$ measures the total 4-momentum in a given energy distribution. 

Any observable is defined by the way it acts on a set of final state
particles. We consider observables whose value only depends on the
momenta of the particles, and not on any additional quantum
numbers. Such observables can be defined in terms of a fully
differential energy distribution, following the discussion of
\cite{Bauer:2008dt,Bauer:2008jx}.

We introduce the phase space $\Phi_B$ of the underlying Born
configuration, that in our case is defined by the two back-to-back
quarks $p$ and $\bar{p}$ prior to any emission. It is constructed in
terms of the energy distribution through the total momentum
$P^\mu[\w^{(N)}] = Q(1,\vec{0})$ and the thrust axis $\vec{n}$.
The cumulative distribution for a given observable can be constructed
from the projection of the cross section that is fully differential in
the energy distribution of the event, onto the value of the
observable. We express $\Sigma(v)$ as~\footnote{The integration over
  $\Phi_B$ is trivial in the process considered in this
  article. However, in the more general case one should integrate over
  all possible Born configurations.}
\begin{equation}
\Sigma(v) \equiv \int \!\df \Phi_B \,\Sigma(\Phi_B; v)\,,
\end{equation}
and define 
\begin{align}
\label{eq:general_obs_def}
  \Sigma(\Phi_B; v)
  &\equiv  \int\!\Df\w\,\frac{\delta\sigma}{ \delta\w}  \,
    \Theta\left(V[\w] < v\right)\delta(\Phi_B - \Phi_B[\w])
    \nn
  &\equiv \sum_N \int\!\Df\w^{(N)}\,\frac{\delta\sigma}{ \delta\w^{(N)}}  \, \Theta\left(V[\w^{(N)}] < v\right) \delta\left(\Phi_B - \Phi_B[\w^{(N)}]\right)
    \,.\end{align}
  Here $\Phi_B[\w^{(N)}]$ denotes the direction of the initial (Born) quarks
  that can be entirely reconstructed from the final-state emissions
  $k_i$ and from the final  (i.e. after all radiation
  has been emitted) quark momenta $q_n$, $q_\bn$.
  The integration measure $\Df\w^{(N)}$ was defined in \eq{Dwn}, and
  the observable for any fixed $N$ can again be written in terms of
  the momenta $k_i$, $q_n$ and $q_\bn$
\begin{align}
V[\w^{(N)}]\equiv V[q_n, q_\bn, k_1, \ldots, k_N] 
\,.
\end{align}
The fully differential cross section is given
by the square of the matrix element as
\begin{align}
\label{eq:fullyDifferential_M}
\frac{\delta\sigma}{\delta\w^{(N)}} \equiv \frac{1}{2F}
\left| M(q_n, q_\bn; k_1, \ldots, k_N)\right|^2 
\,,
\end{align}
where $F$ denotes the flux factor, and $| M(q_n, q_\bn; k_1, \ldots, k_N)|^2$ the squared matrix elements to all orders in perturbation theory with a fixed number $N$ of  emissions. 

We conclude the notation section by giving our definition of the running coupling constant that will be used throughout the paper. The dependence of $\alpha_s$ on the renormalization scale is given in general by
\begin{align}
\mu_R^2 \frac{\df}{\df \mu_R^2} \alpha_s(\mu_R) = -  \alpha_s(\mu_R) \left[ \beta_0 \frac{\alpha_s(\mu_R)}{4\pi} + \beta_1 \left(\frac{\alpha_s(\mu_R)}{4\pi}\right)^2 + \ldots \right]
\,,\end{align}
where 
\begin{align}
\beta_0 = \frac{11 \,C_A-2 n_f}{3}\,, \qquad \beta_1 = \frac{34 \, C_A^2 - 10 \, C_A n_f-6\, C_F n_f}{3}
\,.\end{align}

\section{Basic idea of numerical resummation for rIRC safe observables}
\label{sec:basic_idea}
To resum a cumulative distribution of a given observable in SCET, one
typically proceeds in various steps. First, one needs to identify the
degrees of freedom of the effective theory, which allows one to
reproduce the singular behavior of the observable under
consideration. Typically, this requires a combination of collinear
modes along the directions of the hard particles of the underlying
Born process, and soft modes, which can provide interactions between
different collinear directions. The scaling of the collinear and soft
modes depends on the observable. For example, for thrust ($\tau = 1-T$)
the scaling is
\begin{align}
\sqrt{p_s^2} \sim Q \tau\,, \qquad \sqrt{p_J^2} \sim Q \sqrt{\tau}
\,,
\end{align}
for soft and collinear modes, respectively. These modes are then used
to define the effective SCET
Lagrangian\cite{Bauer:2000ew,Bauer:2000yr,Bauer:2001ct,Bauer:2001yt}. Resummation
is then normally achieved by factorizing the cumulative distribution
into a hard matching coefficient, together with soft and collinear
(jet) functions, and then solving RG equations for the various
components of the factorization theorem. The details of this
factorization theorem as well as that of the RG equations depend again
on the observable.

As stated in the introduction, in this paper we approach the
resummation from a different angle. Within the framework of SCET, one
proceeds as follows:
\begin{enumerate}
\item Given an observable $V(q_n, q_\bn, k_1, \ldots, k_N)$, build
  a simplified version of the observable
  $V_{\rm max}(q_n, q_\bn, k_1, \ldots, k_N)$, which shares the same
  leading logarithmic structure with $V$.

\item Identify the degrees of freedom for the simplified observable
  $V_{\rm max}$, and build the corresponding SCET Lagrangian. Use this
  Lagrangian to obtain a factorization formula for the fully
  differential energy distribution used in \eq{general_obs_def}.

\item Decompose the resummed cumulative distribution as
\begin{equation}
\Sigma(\Phi_B ;v) = \Sigma_{\rm max}(\Phi_B ;v) {\cal F}(\Phi_B ;v)\,,
\end{equation}
where $\Sigma_{\rm max}$ is the cumulative distribution for
$V_{\rm max}$, while ${\cal F}$ is a {\it transfer function} that
relates the latter to the full observable.

\item The resummation is achieved by performing the resummation for
  $\Sigma_{\rm max}(\Phi_B ;v)$ analitically using the usual SCET technology, and then
  computing the corresponding transfer function ${\cal F}(\Phi_B ;v)$
  using the above Lagrangian.
\end{enumerate}

As we will show, the factorization theorem for the simple observable
in SCET, and hence its resummation, is straightforward and can be
obtained in a general manner. The feature that makes this approach
powerful is that the transfer function can be computed numerically for
wide classes of observables. As a consequence, this technique can be
used to resum observables without the need for an observable-dependent
factorisation theorem, as it is usually done in SCET.

In the following sections we will go through the derivation of each of
the above steps in some more detail.

\subsection{Factorization of the fully differential energy distribution}
The starting point for our formulation is the expression for the
cumulative distribution given in \eq{general_obs_def}.  As was shown
in~\cite{Bauer:2008jx}, the
 fully differential energy distribution
$\delta\sigma / \delta\w$ can be written in terms of the
energy momentum tensor of SCET. Following the notation
of~\cite{Bauer:2008jx}, we define
\begin{equation}
\label{eq:Eflow_action}
\cE^0(\W) \ket{X} = \w_X(\W) \ket{X}
\,,\end{equation}
where $X$ denotes a general state. 
The energy flow operator $\cE^0(\W)$ can be defined in terms of the energy-momentum tensor
\begin{equation}
T^{\mu\nu}
= \sum_{\phi\in\cL} \frac{\partial\cL}{\partial(\partial_\mu\phi)}\, \partial^\nu \phi - g^{\mu\nu}\cL
\,,\end{equation}
as~\cite{Korchemsky:1997sy, Bauer:2008dt}
\begin{equation}
\label{eq:Eflow_def}
\cE^0(\W) = \lim_{R\to \infty} R^2 \int_0^\infty\!\df t \, \vn^i\, T^{0 i}(t, R\, \vn)
\,.\end{equation}
Here, $\vn \equiv \vn(\W)$ is the unit three-vector pointing in the direction identified by $\W$. Therefore, $\cE^0(\W)$ measures the total energy arriving over time at infinity in the direction $\W$. An explicit proof of \eq{Eflow_def} for scalars and Dirac fermions can be found in Ref.~\cite{Bauer:2008dt}.

As it was shown in~\cite{Bauer:2008jx}, given that the energy momentum tensor is linear in the Lagrangian of the theory, it can be written as
\begin{equation} \label{eq:Eflowlinearity}
\cE^0(\W) = \sum_{\ell} \cE^0_{n_\ell}(\W) + \cE^0_s(\W)
\,,\end{equation}
where $\cE^0_{n_\ell,s}(\W)$ are defined analogously to
\eq{Eflow_def}, but using the energy-momentum tensor obtained from the
Lagrangian $\cL_{\ell,s}$ only. This then implies that the fully
differential energy distribution can be written in the factorized
form~\cite{Bauer:2008dt,Bauer:2008jx}
\begin{align} 
\label{eq:dsigma_dw}
\frac{\delta \sigma}{\delta \w}
&= \abs{C(\Phi_B)}^2\int\!\Df \w_n  \, \frac{\delta \sigma_n}{\delta \w_n}  \int\!\Df \w_\bn  \, \frac{\delta \sigma_\bn}{\delta \w_\bn} \int\!\Df\w_s\, \frac{\delta \sigma_s}{\delta \w_s}
 \delta\bigl(\w - \w_s - \w_n - \w_\bn\bigr)
\,.\end{align}
Here $ \abs{C(\Phi_B)}^2$ denotes the matching coefficient describing
the short distance fluctuations in the full theory that are not
included in SCET, which depends on the underlying Born configuration
of the process under consideration, but is independent of the
definition of the observable. The terms $\delta \sigma_F / 
\delta \w_F$ denote the fully differential cross section as computed
from the part of the SCET Lagrangian describing sector $F \in \{s, n, \bn\}$. 
In each sector, one can write
\begin{align}
\label{eq:Sigma_general_sector} 
\frac{\delta\sigma_F}{\delta\w_F}
&\equiv \sum_N \frac{\delta\sigma_F}{ \delta\w_F^{(N)}} 
\,.\end{align}
In the soft sector, one finds
\begin{align}
\label{eq:fullyDifferential_M_soft}
\frac{\delta\sigma_s}{\delta\w_s^{(N)}} \equiv \Bigg\{ \begin{array}{l} 
N = 0 : \quad {\cal V}_s \,\delta(\omega_s) \\
N > 0: \quad  {\cal V}_s \,\left| M_s(k_1, \ldots, k_N)\right|^2 
\,,
\end{array}
\end{align}
where the virtual corrections ${\cal V}_s$ and the real emission matrix
element are computed using the Feynman rules of the given sector.  The
only difference with \eq{fullyDifferential_M} is the absence of the
differential Born cross section, which is contained in the matching
coefficient $\abs{C(\Phi_B)}^2$. In the collinear one instead has
\begin{align}
\label{eq:fullyDifferential_M_collinear}
\frac{\delta\sigma_n}{\delta\w_n^{(N)}} \equiv \Bigg\{ \begin{array}{l} 
N = 0 : \quad {\cal V}_n \,\delta(\omega_n-\w[q_n]) \\
N > 0: \quad  {\cal V}_n \left| M_n(q_n; k_1, \ldots, k_N)\right|^2 
\,.
\end{array}
\end{align}
An important difference between the collinear and the soft sector is
that in the latter there is still a single quark $q_n$ contributing to
the energy distribution ($\w_n = \w[q_n]$), even in the absence of any
radiation.

Combining \eq{dsigma_dw} with \eq{general_obs_def} one can write\begin{align}
\label{eq:Sigma_general_fac}
\Sigma(\Phi_B; v) = & \abs{C(\Phi_B)}^2\int \Df \w \int\!\Df \w_n  \, \frac{\delta \sigma_n}{ \delta \w_n}  \int\!\Df \w_\bn  \, \frac{\delta \sigma_\bn}{ \delta \w_\bn} \int\!\Df\w_s\, \frac{\delta \sigma_s}{ \delta \w_s}
\nn & \quad \times
\Theta(V[\w] < v) \, \delta\bigl(\w - \w_s - \w_n - \w_\bn\bigr)
\,,
\end{align}
where the sum of energy densities is obtained by taking the union of
the momenta in each sector, as defined in \eq{omega_union}. 

Note that \eq{Sigma_general_fac} does not imply a factorization formula for $\Sigma(\Phi_B; v)$ in the commonly used sense, since the observable $V[ \w_n + \w_\bn + \w_s]$ does not factorize in general. Only observables which do not combine momenta from different sectors in a non-trivial way, such that they can be written as 
\begin{align}
\label{eq:obs_fact}
V[\w_n + \w_\bn + \w_s] = G[V[\w_n], V[\w_\bn],  V[\w_s]]
\,,\end{align}
satisfy a commonly used factorization theorem. 
One then obtains
\begin{align}
\Sigma(\Phi_B; v) = & \abs{C(\Phi_B)}^2\int\!\df v_n  \, \frac{d \sigma_n}{ \df v_n}  \int\!\df v_\bn  \, \frac{\df \sigma_\bn}{ \df v_\bn} \int\!\df v_s\, \frac{\df \sigma_s}{ \df v_s}
\Theta(G[v_n, v_\bn, v_s] < v)
\,,\end{align}
where
\begin{align}
\frac{d \sigma_F}{ \df v_F} = \int \!\Df \w_F \, \frac{\delta \sigma_F}{ \delta \w_F} \, \delta\left(V[\w_F] - v_F \right)
\,.
\end{align}
Thus, for factorizable observables one reproduces the factorized result that was the starting point in~\cite{Bauer:2018svx} when discussing the thrust distribution at NLL accuracy. 
For example, an additive observable, such as thrust satisfies
\begin{align}
V[ \w_n + \w_\bn + \w_s] = V[ \w_n] + V[ \w_\bn] + V[ \w_s]
\,.
\end{align}

Another important consequence of \eq{Sigma_general_fac} is that, in
general, the delta function constraining the total energy density to
the sum of the ones in each sector introduces a kinematic cross-talk
between the soft and collinear sectors. This can be understood by
noting that
\begin{align}
\label{eq:energy_distribution_conservation}
P^\mu[\w] = P^\mu[\w_s]  + P^\mu[\w_n] + P^\mu[\w_\bn] = Q(1, \vec{0})
\,.
\end{align} 
Thus, the quark that is initiating each collinear sector recoils
against the soft radiation in the corresponding hemisphere. This is
important for observables that are sensitive to the recoil of the Born
quarks, such as the jet broadening. 


\subsection{Definition of the simple observable}
\label{sec:simple_observable_def}
The next essential step is the definition of the simple observable
$V_{\rm max}$. The only constraint on such an observable is that it
needs to have the same leading logarithmic structure as the full
observable $V$. Many choices are of course possible, but it is
convenient to adopt a simple definition that is generic and can be
easily handled analytically.

A particularly simple and systematic procedure is to define a global
and rIRC safe observable in each sector individually, and then to
compute the final value of the simple observable as the maximum of the
value in each sector. In other words, we define
\begin{align}
\label{eq:measurement_function_Vmax}
  V_{\rm max}[\w_n + \w_\bn + \w_s] = {\rm max}[V_{\rm max}[\w_n], V_{\rm max}[\w_\bn],  V_{\rm max}[\w_s]] \equiv {\max}[v_n, v_\bn, v_s ]\,,
\end{align}
where the definition of $V_{\rm max}[\w_F]$ in each sector $F$ is
given below.

\paragraph{The soft sector\\}
Since the observable needs to be defined in resummed perturbation
theory, one needs to define it for an arbitrary number of final state
soft particles. We decompose the squared amplitude
$\left|M_s(k_1, \ldots, k_N)\right|^2$ for producing a set of
particles with momenta $k_1$ to $k_N$ into soft webs, that in the
following will be referred to as correlated clusters\footnote{Note
  that the correlated clusters $\tilde M_s^2$ are in general not
  positive definite.}
\begin{align}
\nonumber
\left|M_s(k_1)\right|^2 &\equiv \tilde{M}_s^2(k_1) \\\nonumber 
\left|M_s(k_1, k_2)\right|^2 &= \tilde{M}_s^2(k_1) \tilde{M}_s^2(k_2) + \tilde{M}_s^2(k_1,k_2) \\ 
\left|M_s(k_1, k_2, k_3)\right|^2 &= \tilde{M}_s^2(k_1) \tilde{M}_s^2(k_2)
                     \tilde{M}_s^2(k_3)\nn
& \qquad + \left(\tilde{M}_s^2(k_1)
                     \tilde{M}_s^2(k_2,k_3) + \text{perm.}\right)
+
                     \tilde{M}_s^2(k_1,k_2,k_3)\, \nonumber\\
\label{eq:clusters}
\vdots 
\end{align}
Each cluster is recursively defined as the portion of the squared
amplitude that can not be written in terms of products of
lower-multiplicity clusters. The 1-particle correlated cluster
$\tilde{M}_s^2(k)$ describes the square of the eikonal amplitude for
emitting a single gluon from the Wilson line. For the emission of two
gluons, the correlated cluster $\tilde{M}_s^2(k_1,k_2)$ is defined as
the difference of the 2-gluon squared eikonal amplitude
$\left|M_s(k_1, k_2)\right|^2$ and the product of two 1-gluon squared
eikonal amplitudes determined in the previous step. For the emission
of a quark anti-quark pair, the correlated cluster is given by the
full $q \bar q$ eikonal squared amplitude, since the 1-particle
eikonal squared amplitude with only a quark does not exist. The
extension to higher multiplicities proceeds in a straightforward
fashion. Each correlated cluster admits a perturbative expansion
arising from the virtual corrections, and we write
\begin{align}
\tilde{M}_F^2(k_1, \ldots,  k_N) = \sum_{i=0}^\infty \left(\frac{\alpha_s}{2\pi}\right)^i \tilde{M}_{F,i}^2(k_1, \ldots,  k_N) 
\,.
\label{eq:cluster_expansion}
\end{align}

Using this structure, one can define the simple observable
$V_{\rm max}$ in the soft sector through its action on products of
correlated clusters
\begin{align}
\label{eq:VMax_n}
&\tilde M_s^2(k_1, \ldots, k_{m_1}) \ldots \tilde M_s^2( k_{m_k+1}, \ldots, k_{N}) \,V_{\rm max}(q_n, q_\bn; k_1,\dots,k_N)
\\
& \qquad = \tilde M_s^2(k_1, \ldots, k_{m_1}) \ldots\tilde M_s^2(k_{m_k+1}, \ldots, k_{N}) 
\nn
& \qquad \qquad \times \max\{\tilde V_s(k_1 + \ldots + k_{m_1}), \dots, \tilde V_s(k_{m_k+1}+ \ldots+ k_{N})\}
\,.
\end{align}
One therefore only needs a definition of $\vtS(k)$, acting on the
total momentum of each correlated cluster. 

A convenient choice is to consider the following generalisation of the moments of
energy-energy correlation~\cite{Banfi:2004yd}
\begin{equation}
\label{eq:FC_ab}
{\rm FC}^{(a,b)} = \sum_{i\neq j} \frac{E_i^a
  E_j^a}{Q^{2 a}}
\left|\sin\theta_{ij}\right|^{a-b}(1-\left|\cos\theta_{ij}\right|)^{b}\,,
\end{equation}
and to define the simple observable using the following procedure:

\begin{enumerate}
\item Associate each momentum $k_i$ (for correlated clusters with more
  than one momentum, $k_i$ corresponds to the total momentum of the
  correlated cluster) with the Wilson lines it is closest in rapidity.
  This effectively divides the event into two hemispheres.

\item For each hemisphere, compute the contribution of each correlated
  cluster to eq.~\eqref{eq:FC_ab} in the {\it soft-collinear}
  limit. 
  If the correlated cluster contains more than one particle, i.e. it
  has a non-vanishing invariant mass, we evaluate eq.~\eqref{eq:FC_ab}
  in the massless approximation, i.e. we change the energy of the
  cluster $k^0$ such that $k^2=0$.
  This amounts to considering the energy-energy correlator
  between each of the clusters and the quark moving in the same
  hemisphere (after all soft radiation has occurred), that can be
  recast as
\begin{align}
\label{eq:Vmax_0}
\vtS(k)  = \left(\frac{k_\perp}{Q}\right)^{a} e^{-b_\ell \eta_\ell}\,,
\end{align}
where $k_\perp$ and $\eta_\ell$ are the transverse momentum and
pseudo-rapidity of the considered correlated cluster with respect to the
above quark, labelled by $\ell=\{n, \bn\}$.\footnote{Most observables
  satisfy $b_n = b_\bn = b$. However, our considerations can be easily
  generalized to cases with different $b_\ell$ in each collinear
  direction. In this case one can simply consider different
  moments~\eqref{eq:FC_ab} in each collinear sector.}
The final momentum of the two quarks $\tilde{q}_n$, $\tilde{q}_\bn$
after the emission of the soft radiation (note that these differ from
the final state quarks $q_n$ and $q_\bn$ by the recoil associated with
collinear emissions) is simply obtained by the Born momenta as
\begin{align}
\tilde{q}_n^\mu  = p^\mu - \sum_{i \in {\cal H}_n} k_i^\mu  \,, \qquad \tilde{q}_\bn^\mu = \bar{p}^\mu - \sum_{i \in {\cal H}_\bn} k_i^\mu
\,,
\end{align}
where $p^\mu$ and $\bar{p}^\mu $ are the Born momenta prior to any
emissions.
The directions of the two momenta $\tilde{q}_n$ and $\tilde{q}_\bn$ then define the
direction of the Wilson lines used in the (recoil-free) definition of the SCET
Lagrangian.

The parameters $a$ and $b_\ell$ must be chosen in order to match the {\it
  soft-collinear} scaling of the full observable $V$, which guarantees
that the simple observable has the same leading logarithmic structure
as the full observable.

\item Finally, define the simple observable $v_s \equiv V_{\rm
    max}[\omega_s]$ as the largest of all of the $\vtS(k)$ values.
\end{enumerate}

The above definition corresponds to a toy observable that will lead to
important simplifications both in its analytical resummation and in
the numerical evaluation of the transfer function.

\paragraph{The collinear sector\\}
As will be discussed later, the collinear sector only contributes
non-trivially to NNLL accuracy and beyond. Given the discussion in
Section~\ref{sec:logcounting}, this implies that in the contribution
of the collinear sector to the transfer function one needs to consider
only a fixed number of emissions at a given logarithmic order, with
$r$ emissions needed to obtain N$^{r+1}$LL accuracy.  Thus, to obtain
NNLL accuracy a single collinear emission suffices, while at N$^3$LL
accuracy at most two collinear emissions are required. For this reason, a
decomposition of the collinear squared amplitudes in terms of
correlated clusters is not necessary, and we can rather define the
simple observable in terms of its action on the full collinear squared
amplitude. In each of the collinear sectors, say $n$, the simple
observable $V_{\rm max}[\w_n]$ is then simply given by
eq.~\eqref{eq:FC_ab} in the small $\theta_{ij}$ approximation,
namely
\begin{align}
\label{eq:Vmax_coll}
v_n\equiv V_{\rm max}[\w_n]  = \tilde{V}_{n}(k_1,\dots) = \sum_{i\neq j} \frac{E_i^a
  E_j^a}{Q^{2 a}}
\left|\theta_{ij}\right|^{a+b_n} 2^{-b_n}\,,
\end{align}
where the sum runs over all particles belonging to the collinear
sector $n$.
For a single emission, a useful parametrisation of
$v_n\equiv \tilde{V}_{n}(k_n)$ is given in eq.~\eqref{eq:Vtilde_n}.

We note that this simplified observable in the collinear sector only
depends on the momenta in the final state, but is independent of the
axis used to define the collinear Lagrangian. The corresponding jet
function is therefore independent of any recoil against the soft sector.

\subsection{The SCET Lagrangian and resummation of the simple observable}
From the above definition of the simple observable, it is now clear
that the scaling of the soft and collinear modes in SCET can be
obtained in terms of the $a$ and $b_\ell$ parameters as
\begin{align}
\label{eq:SCET_scaling}
\sqrt{p_s^2} \sim Q v^{1/a}\,, \qquad \sqrt{p_{J_\ell}^2} \sim Q v^{1/(a+b_\ell)}
\,.
\end{align}
The SCET Lagrangian is uniquely defined by the modes given above
\begin{align}
{\cal L}_{\rm SCET} = {\cal L}_s + \sum_{\ell \in \{n,\bn\}} {\cal L}_{\ell}
\,.
\end{align}

The measurement function~\eqref{eq:measurement_function_Vmax} allows
for a simple factorization formula in the traditional sense. Using
\begin{align}
\Theta({\max}[v_n, v_\bn, v_s ] < v) = \Theta(v_n< v )\, \Theta(v_\bn< v )\, \Theta(v_s< v)
\,,
\end{align}
and the fact that the simple observable is recoil
insensitive~\cite{Larkoski:2014uqa}, one finds
\begin{align}
\label{eq:Sigmamax_fact}
\Sigma_{\max}(\Phi_B; v) = & \abs{C(\Phi_B)}^2\, \Sigma_n^{\max}(v) \,\Sigma_\bn^{\max}(v) \,\Sigma_s^{\max}(v)
\,,
\end{align}
where the expression in each sector is given by the obvious result
\begin{align}
\label{eq:Sigma_max_F_def}
\Sigma_F^{\max}(v) = \int\!\Df\w_F\,\frac{\delta\sigma_F}{ \delta\w_F} \,\Theta(V_{\rm max}[\w_F] < v)
\,.
\end{align}
Eq.~\eqref{eq:Sigmamax_fact} can be handled with standard RG
techniques as it is usually done in SCET, as discussed in more detail
in Appendix~\ref{app:Sigma_max_details}.

From now on we restrict ourselves to observables whose infrared
dynamics is described by the degrees of freedom of this SCET
Lagrangian.
It is well known that more involved observables may require additional
modes. An example is given by joint
resummations~\cite{Bauer:2011uc,Larkoski:2014tva,Procura:2014cba}, or
by some cases of single resummations (such as negative-$b$
angularities defined w.r.t. the thrust axis\footnote{See for instance
  the discussion in Appendix I.1 of ref.~\cite{Banfi:2004yd}.}) where
some extra care must be taken when constructing the effective
Lagrangian. We believe that the formulation presented in this article
can be adapted to such cases, although we will not discuss this
further in the present article.

\subsection{Definition of the transfer function}
The second ingredient for the resummation is the transfer function
${\cal F}$, which relates the resummation of the simple observable
$V_{\rm max}$ to that of the desired observable $V$. Most global rIRC
safe observables have the property that the cumulative distribution is
exponentially suppressed [e.g.
$\Sigma(v) \sim \exp(-\alpha_s \ln^2 v)$] in the limit $v \to 0$. Such
observables satisfy the basic relation to the cumulative distribution
of their simple observables~\cite{Bauer:2018svx}\footnote{Observables
  that involve kinematic cancellations, such as the transverse
  momentum of a colour singlet system at hadron colliders, require
  additional considerations~\cite{Monni:2016ktx}.}
\begin{align}
\label{eq:fullF_def}
\Sigma(\Phi_B; v) = \Sigma_{\rm max}(\Phi_B; v)\, \fullF(\Phi_B; v)
\,.
\end{align}
Using \eqs{Sigma_general_fac}{Sigmamax_fact} one can write a general factorization theorem for the transfer function
\begin{align}
\label{eq:transfer-nonfact}
\fullF(\Phi_B; v) 
=  & \int\!\Df \w_n  \, \fullF'_n(\w_n, v)
\int\!\Df \w_\bn  \,  \fullF'_\bn(\w_\bn, v)
\int\!\Df\w_s\,  \fullF'_s( \w_s, v)
\nn & \quad \times
\Theta(V[ \w_n + \w_\bn + \w_s] < v)
\nn
\equiv  & \int\!\Df \w  \, \fullF'(\Phi_B; \w, v) \, \Theta(V[ \w] < v)
\,,
\end{align}
where
\begin{align}
\label{eq:fullF_conv_def}
\fullF'(\Phi_B; \w, v) \equiv & \int\!\Df \w_n  \, \fullF'_n( \w_n, v)
\int\!\Df \w_\bn  \,  \fullF'_\bn( \w_\bn, v)
\int\!\Df\w_s\,  \fullF'_s( \w_s, v)
\nn & \quad \times
\delta(\w - \w_s - \w_n - \w_\bn)\,,
\end{align}
and the individual fully differential transfer functions are given by
\begin{align}
\label{eq:fullyDiffTransfer_def}
\fullF'_F( \w_F, v) = \frac{\frac{\delta \sigma_F}{ \delta \w_F}}{\Sigma_F^{\max}( v)} 
\,.
\end{align}
Note that the form of the general factorization theorem given in
\eq{transfer-nonfact} {\it does not} depend on the observable, and in
particular it {\it does not} depend on whether the observable
factorizes or not.

A couple of considerations are in order.
In the transfer functions for each separate sector, the virtual
contributions to the numerator and denominator cancel in the ratio,
such that only real emissions contribute. As explained in
\cite{Bauer:2018svx}, and shown in the following sections,
\eq{fullyDiffTransfer_def} can be computed numerically provided that
the UV divergences in the SCET real radiation are handled by means of
a regulator different from dimensional regularization, such as a
cutoff or an analytic regulator. In this case the transfer function is
finite in 4 dimensions and can be evaluated efficiently.

\section{General expressions for the fully differential transfer function}
\label{sec:transfer_fully_differential}

In this section we derive the fully differential transfer function at
NLL and NNLL accuracy. In our derivation, we only use the general
factorization theorem given in~\eq{dsigma_dw} that is valid for any
observable that depends only on the kinematics of the final state, but
without assuming that the measurement function of the desired
observable itself factorizes. This therefore extends the results
of~\cite{Bauer:2018svx}, in which similar results were derived for the
thrust distribution, to generic observables.
Note that in order to keep this section as simple to understand as
possible, we write the results in terms of general squared amplitudes
of the effective theory, and we defer giving the results using the
explicit expressions for them obtained in SCET to
Section~\ref{sec:explicit_transfer}.

In order to obtain an expression that is suitable for a numerical
integration, one needs to address the cancellation of both IRC and UV
divergences. We discuss them separately in the following.

\subsection{Numerical treatment of the UV divergences}
\label{sec:numerical_treatment_UV_regulator}

As already discussed in ref.~\cite{Bauer:2018svx}, UV divergences
appear both in the real and in the virtual corrections in a SCET
calculation. In SCET$_{\rm I}$ theories, where soft and collinear
degrees of freedom have different virtualities, both of these UV
divergences are regulated by dimensional regularization, which then in
turn leads to the anomalous dimensions of the soft and jet
functions. On the other hand, in SCET$_{\rm II}$ theories, where soft
and collinear degrees of freedom have the same virtuality, an
additional regulator must be introduced to regularize the rapidity UV
divergences that arise from the real radiation. Possible choices
involve different analytic
regulators~\cite{Chiu:2009yx,Becher:2011dz,Chiu:2012ir,Echevarria:2015byo,Li:2016axz}.

Since in the numerical approach to resummation one needs to be able to
perform real phase space integration numerically in 4 dimensions
(after IRC divergences have been properly subtracted), the UV
divergences in SCET$_{\rm I}$ theories need to be regulated with a
regulator other than dimensional regularisation. This guarantees that
real and virtual UV divergences are regulated independently, and thus
can be handled separately.
The exponential regulator introduced in~\cite{Li:2016axz} regulates
the UV rapidity divergence by shifting the separation of the soft
Wilson lines that define the soft operator, and the collinear fields
that define the collinear operator, by a small complex amount
$\tau \sim 1 / \Lambda$ in both the $x^\pm$ light-cone directions. In
momentum space, this gives rise to an exponential suppression of the
form
\begin{align}
\label{eq:transfer_NkLL exp_regulator}
\exp\left[ -2 e^{-\gamma_E}k_0 / \Lambda\right]
\,,
\end{align}
where $k_0$ denotes the total energy that flows across the real
phase-space cut (the total energy of the real radiation). 
Effectively, this amounts to regulating the UV divergences at a scale
$\sim \Lambda$. Such divergences will then appear as logarithms of
$\Lambda$.
Since the rapidity divergences cancel between the soft and collinear
sectors, the dependence on $\ln \Lambda$ vanishes once these sectors
are combined, but it does change the RG equations of the soft and
collinear sector individually.

To isolate the logarithmic dependence on $\Lambda$ which characterizes
these extra UV divergences (and neglect any power suppressed
dependence), one needs to expand the integrals obtained in SCET around
the limit $\Lambda \to \infty$. In a numerical implementation such an
expansion is more difficult to do. One could simply choose a very
large numerical value of $\Lambda$, such that the power suppressed
effects would be numerically small, but this can result in an
inefficient numerical algorithm. A better approach is to take formally
the limit $\Lambda \to \infty$ of the exponential suppression factor
{\it before} performing the integration, such that it only gives rise
to the logarithmic terms upon integration.
For the soft integrals this amounts to making the asymptotic
replacement~\footnote{This identity can be easily verified numerically
  by taking the large-$\Lambda$ limit of
  eq.~\eqref{eq:transfer_NkLL exp_regulator}.} 
\begin{align}
\label{eq:rapidity-regulator-limit}
\exp\left[ -2 e^{-\gamma_E} k_0 / \Lambda\right] \stackrel{\rm soft}{\to}  \Theta\left(k^- < k^+\right) \Theta_{\rm soft}[k^+, \Lambda] + \Theta\left(k^+ < k^-\right) \Theta_{\rm soft}[k^-, \Lambda]
\,,\end{align}
where
\begin{align}
  \Theta_{\rm soft}[k, \Lambda] \equiv \Theta\left(k < \Lambda\right)
  \,. \end{align}
This converts the regulator into a cutoff on the larger light-cone
component of each real momentum. One restricts the momentum $k^+$ in
the integration region $k^+ > k^-$ and or $k^-$ for $k^- > k^+$. 
For the collinear integrals, the purely collinear contributions are
already regulated in the UV by the phase space limits, and therefore
the exponential regulator does not give rise to any logarithmic
contributions and it can be entirely dropped. On the other hand, for
the 0-bin subtraction in the collinear sector in the $n$ direction,
one proceeds as in the soft sector and uses the constraint
$\Theta_{\rm soft}[\bn \cdot k, \Lambda]$ for each emission. The above
prescription amounts to defining the 0-bin subtraction integrals by
replacing the collinear squared amplitude by 
\begin{align}
\label{eq:rapidity-regulator-limit-coll}
\left|M_{n}(k_1,\dots, k_N)\right|^2 \to  \left|M_{n,0-{\rm bin}
}(k_1,\dots, k_N)\right|^2 \prod_{i=1}^N\Theta_{\rm soft}[\bn \cdot k_i, \Lambda]
\,.\end{align}

In practical calculations this reproduces precisely the cutoff
prescription discussed in~\cite{Bauer:2018svx}, while retaining the
nice properties of the exponential regulator in the analytic
formulation. Since the dependence on $\Lambda$ cancels between the
soft and collinear sectors, one can choose any value of the UV scale
for the practical numerical calculation. As discussed in
ref.~\cite{Bauer:2018svx}, a particularly convenient choice is to set
it to its characteristic scale of the collinear sector, i.e.
$\Lambda = Q$. This choice makes the collinear sector single
logarithmic, while all double logarithms are included in the soft
sector of the SCET Lagrangian. In the following we adopt this scheme
in the derivation of the expressions for the transfer function.

Since the collinear sector is single logarithmic, the fully
differential collinear transfer function at NLL reduces to a $\delta$
function, giving
\begin{align}
  \fullF^{'\rm NLL}_{n, \bn}( \w_{n,\bn}, v) = \delta(\w_{n,\bn} - \w[q_{n,\bn}]) 
  \,,\end{align}
such that it only depends on the energy density introduced by the
single recoiling quark of the underlying Born
configuration. Conversely, the soft transfer function at NLL requires an infinite number of 1-particle soft clusters at tree-level.

At NNLL, the collinear transfer function requires a single collinear emission at tree level, while for the soft transfer function one has to add a single insertion of the 1-loop contribution to the soft 1-particle cluster, and of the soft 2-particle cluster at tree-level. This generalizes in a systematic way to any logarithmic order. 
The transfer function at a general order N${}^k$LL requires a certain
number of insertions of $n$-particle soft clusters with
$n = n_{\rm particles}$ computed to $n_{\rm loop}$ order [see
\eq{cluster_expansion}]. Given the NLL transfer functions, described
above, the general expression for $k > 1$ can be written recursively
as
\begin{align}
\fullF^{'\rm N^kLL}_F( \w_F, v) = \fullF^{'\rm N^{k-1}LL}_F( \w_F, v) + \delta\fullF^{'\rm N^kLL}_F( \w_F, v)\,.
\end{align}
The insertions needed the computation of the corrections
$\delta\fullF^{'\rm N^kLL}_F( \w_F, v)$ are summarized in
Table~\ref{tab:LogCounting}.
\begin{table}
\begin{center}
\begin{tabular}{|c||c|c||c|c|}
\hline 
& \multicolumn{2}{c ||}{soft sector (clusters)} & \multicolumn{2}{c |}{collinear sector}
\\\hline
& $n_{\rm loops} + n_{\rm particles}$ & times & $n_{\rm loops} + n_{\rm particles}$ & times
\\ \hline\hline
& \multicolumn{2}{c ||}{$\fullF'_s$} & \multicolumn{2}{c |}{$\fullF'_{n, \bn}$}
\\\hline
NLL & $1$ & $\infty$ & -- & --
\\ \hline\hline
& \multicolumn{2}{c ||}{$\delta \fullF'_s$} & \multicolumn{2}{c |}{$\delta \fullF'_{n, \bn}$}
\\\hline
NNLL & $2$ & $1$ & $1$ & $1$
\\ \hline
N${}^3$LL & $3$ & $1$ & &\\
                 & $2$ & $2$ & $2$ & $1$ 
\\ \hline
N${}^k$LL & $k$ & $1$ & & \\
                 & $k-1$ & $2$ & $k-1$ & $1$ \\
                 & $\vdots$ & $\vdots$ & $\vdots$ & $\vdots$ \\
                 & $2$ & $k-1$ & $2$& $k-2$\\\hline 
\end{tabular}
\end{center}
\caption{Ingredients needed for the NLL transfer function, as well as
  for the corrections required to for the the result at N${}^k$LL. For
  each soft and collinear transfer function, the first column of the
  table shows the perturbative accuracy required for the soft
  correlated clusters and the collinear squared amplitudes, while the
  second column shows the number of insertions of each of the above contributions. \label{tab:LogCounting}}
\end{table}

\subsection{Numerical treatment of the IRC divergences}
We start by recalling that, in the definition of the transfer
function~\eqref{eq:fullyDiffTransfer_def},
$\Sigma_{\rm max}(\Phi_B; v)$ is chosen to have the same LL structure
as $\Sigma(\Phi_B; v)$. An important consequence of this fact is that
the pure LL contribution completely cancels in the ratio and the
transfer function differs from one by NLL corrections.

A second important observation is that, as it will be shown shortly,
the contribution from very unresolved emissions entirely cancels in
the ratio when computing the transfer function for a given observable
$V < v$. Therefore, ${\cal F}$ only receives a relevant contribution
from emissions (or correlated clusters in the case of the soft sector)
with $\tilde{V}_F \sim v$. A crucial consequence of this fact is that
the phase space integration associated with each emission loses one
logarithmic power, which in turn implies that the computation of the
transfer function to N$^k$LL accuracy requires the knowledge of the
numerator and denominator of eq.~\eqref{eq:fullyDiffTransfer_def} only
to order N$^{k-1}$LL, that is
\begin{align}
\label{eq:transfer_NkLL}
\fullF^{'\rm N^kLL}_F( \w_F, v) = \frac{\frac{\delta \sigma_F^{{\rm N}^{k-1}{\rm LL}}}{\delta \w_F}}
{\Sigma_F^{{\rm max}, {\rm N}^{k-1}{\rm LL}}( v)} 
\,.
\end{align}

In order to make the cancellation of the IRC divergences explicit in
the transfer function, we introduce a parametrically small quantity
$\eps \ll 1$ which needs to be independent of the observable $v$ in
the $v \to 0$ limit.  This allows us to recast \eq{transfer_NkLL} as
\begin{align}
\label{eq:transfer_NkLL_2}
\fullF^{'\rm N^kLL}_F( \w_F, v) &= \frac{\Sigma_F^{{\rm max}, {\rm N}^{k-1}{\rm LL}}( \eps v)}{\Sigma_F^{{\rm max}, {\rm N}^{k-1}{\rm LL}}( v)}\,
\frac{\frac{\delta \sigma_F^{{\rm N}^{k-1}{\rm LL}}}{\delta \w_F}}
{\Sigma_F^{{\rm max}, {\rm N}^{k-1}{\rm LL}}( \eps v)} 
\nn
& \equiv \Delta_F^{{\rm N}^{k-1}{\rm LL}}( v, \eps v)\,\frac{\frac{\delta \sigma_F^{{\rm N}^{k-1}{\rm LL}}}{\delta \w_F}}
{\Sigma_F^{{\rm max}, {\rm N}^{k-1}{\rm LL}}( \eps v)} 
\,\,.
\end{align}
In this way each of the two factors in the above equation is IRC
finite, and the dependence on the regulator $\eps$ cancels in the
product. 

It is important to notice that the $\Sigma_F^{\rm max}$ factor in
eq.~\eqref{eq:transfer_NkLL_2} is to be computed with the rapidity
regulator defined in the previous section. Although this
regularization procedure can be also adopted in the resummation of the
simple observable in eq.~\eqref{eq:fullF_def}, it is not always
necessary (for instance in the case of SCET$_{\rm I}$ observables) in
the analytic calculation. However, it is crucial that the UV
divergences of the real radiation are regulated as in
Section~\ref{sec:numerical_treatment_UV_regulator} in the computation
of the transfer function~\eqref{eq:transfer_NkLL_2}.
In the following two subsections, we work out the results at
NLL and NNLL in detail.

\subsection{General expression for the transfer function at NLL}
\label{sec:transfer_fully_differential_general_NLL}

According to \eq{transfer_NkLL}, the computation of the transfer
function to NLL accuracy requires the knowledge of
$\delta \sigma_F / \delta \w_F$ and
$\Sigma_F^{\rm max}( v)$ to LL accuracy.  As discussed above,
the collinear sector gives rise to only single logarithmic behavior
such that
\begin{align}
\frac{\delta \sigma_{n,\bn}^{\rm LL}}{ \,\delta \w_{n,\bn}} = \delta\left(\w_{n,\bn} - \w[q_{n,\bn}]\right) \,, \qquad \Sigma^{\rm max, LL}_{n,\bn}( \w_{n,\bn}) = 1
\,,\end{align}
where we have used the fact that $V_{\max}$ is zero for a state that only includes a single quark. 
From this one immediately obtains
\begin{align}
\fullF_{n,\bn}^{'\rm NLL}( \w_{n,\bn}) = \delta\left(\w_{n,\bn} - \w[q_{n,\bn}]\right)
\,.\end{align}
Thus, to NLL accuracy, only the transfer function $\fullF_s( \w_s, v)$ is required and one finds
\begin{align}
\label{eq:transfer_NLL_nonfact}
\fullF^{\rm NLL}(\Phi_B; v) =  & \int\!\Df\w_s\,  \fullF^{'\rm NLL}_s( \w_s, v) \, 
\Theta(V[ \w_s + \w[q_n] + \w[q_\bn]] < v)
\,.
\end{align}

To compute the soft transfer function to NLL accuracy, we start from
the general expression (see eq.~\eqref{eq:Sigma_general_sector}
and~\eqref{eq:fullyDifferential_M_soft})
\begin{align}
\frac{\delta\sigma_s}{ \delta\w_s} = {\cal V}_s \left[ \delta(\w_s) + \sum_{N=1}^\infty \left| M_s(k_1, \ldots, k_N)\right|^2 \right]
\,.
\end{align}
The virtual corrections $ {\cal V}_s$ in the soft sector of SCET are
given by a scaleless integral, and therefore contain both UV and IR
divergences, while with NLL accuracy (as discussed above) we can
approximate
\begin{equation}
\left| M_s(k_1, \ldots, k_N)\right|^2 \simeq \prod_{i=1}^N \tilde M_{s,0}^2(k_i)\,.
\end{equation}
This allows us to write
\begin{align}
\label{eq:numerator_NLL_1}
\frac{\delta\sigma_s^{\rm LL}}{ \delta\w_s} = {\cal V}_s^{\rm LL}  \left[ \delta(\w_s) + \sum_{N=1}^\infty \prod_{i=1}^N \tilde M_{s,0}^2(k_i)\right]
\,,
\end{align}
where ${\cal V}_s^{\rm LL}$ denotes the virtual corrections at LL
accuracy. 
The explicit expression for this soft matrix element will be given in Sec.~\ref{sec:transfer_fully_differential_explicit_NLL}. The precise form of the virtual corrections is not relevant, since it will drop out of the transfer function to all orders. 
The running coupling in each 1-particle correlated cluster needs to be
evaluated at the transverse momentum, which is the only Lorentz
invariant and reparametrization invariant (RPI) scale available in the
soft sector of SCET. $\Sigma_s^{{\rm max}, {\rm LL}} $ in
eq.~\eqref{eq:transfer_NkLL_2} can be written as
\begin{align}
\label{eq:SigmaMaxLL_def}
\Sigma_s^{{\rm max}, {\rm LL}}( v) &= {\cal V}_s^{\rm LL} \sum_{N=0}^\infty \frac{1}{N!} \prod_{i=1}^N \int \! [\df k_i] \tilde M_{s,0}^2(k_i) \, \Theta\left( \vtS (k_i) < v \right)
\nn
&= {\cal V}_s^{\rm LL} \exp \left\{ \int \! [\df k] \tilde M_{s,0}^2(k_i) \, \Theta\left( \vtS (k) < v \right)\right\}
\,,
\end{align}
where $\vtS(k)$ is defined in \eq{Vmax_0}. We have used that the $N=0$
term in the sum is just equal to unity, and that to LL accuracy all
final state particles are gluons, given the symmetry factor
$S(N) = 1 / N!$.

In the combination of \eq{numerator_NLL_1} and \eq{SigmaMaxLL_def} the
virtual correction drops out, and one obtains
\begin{align}
\label{eq:fullF_NLL_1}
\Df\w_s\,  \fullF^{'\rm NLL}_s( \w_s, v)  =  & \Delta_s^{\rm LL}( v, \eps v)
\nn
& \quad \times
\frac{\sum_{N=0}^\infty \frac{1}{N!}\prod_{i=1}^N [\df k_i]  \tilde M_{s,0}^2(k_i)}
{\sum_{N=0}^\infty \frac{1}{N!}\prod_{i=1}^N \int \! [\df k_i]  \tilde M_{s,0}^2(k_i) \, \Theta\left( v_i < \eps v \right)}
\,,
\end{align}
where we have used the shorthand notation $v_i \equiv \vtS(k_i)$, and $\Delta_s^{\rm LL}( v, \eps v)$ has been defined by \eq{transfer_NkLL_2} and can be written as
\begin{equation}
\label{eq:DeltaLL_def}
\Delta_s^{\rm LL}( v, \eps v) \equiv \exp\left[ -\int [\df k]  \tilde M_{s,0}^2(k) \, \Theta\left( \eps v < \vtS(k) < v \right) \right]
\,.
\end{equation}

As a next step, we split the numerator of \eq{fullF_NLL_1} into an unresolved subset (where each $k$ satisfies $\vtS(k) < \eps v$) and a resolved subset (where each $k$ satisfies $\vtS(k) > \eps v$)
\begin{align}
\sum_{N=0}^\infty \frac{1}{N!}\prod_{i=1}^N [\df k_i]  \tilde M_{s,0}^2(k_i) = & \left[ \sum_{N=0}^\infty \frac{1}{N!}\prod_{i=1}^N [\df k_i] \, \Theta( v_i < \eps v) \tilde M_{s,0}^2(k_i) \right]
\nn
& \quad \times \left[\sum_{N=0}^\infty \frac{1}{N!}\prod_{i=1}^N[\df k_i]  \tilde M_{s,0}^2(k_i)\, \Theta( v_i > \eps v)\right]\,.
\end{align}
Now we use that this fully differential transfer function will be
integrated against global rIRC safe observables. They have the
property that unresolved emissions [with $\vtS(k) < \eps v$] can be
neglected up to power corrections in $\eps$ in the measurement
function of the observable~\cite{Banfi:2004yd,Banfi:2014sua}
\begin{align}
& \Theta(\vtS(k_i) < \eps v) \, V(q_n, q_\bn; k_1, \ldots, k_N)
\nn
& \qquad = \Theta(\vtS(k_i) < \eps v) \, V(q_n, q_\bn; k_1, \ldots, k_{i-1}, k_{i+1}, \ldots, k_N) + {\cal O}(v\,\eps^h)
\,,
\end{align}
where $h$ is a positive constant.  Thus, upon integration against such
an observable, and getting rid of power corrections in $\delta$, the
unresolved terms cancel between the numerator and denominator and one
can write
\begin{align}
\frac{\sum_{N=0}^\infty \frac{1}{N!}\prod_{i=1}^N [\df k_i]  \tilde
  M_{s,0}^2(k_i)}{\sum_{N=0}^\infty \frac{1}{N!}\prod_{i=1}^N \int \!
  [\df k_i]  \tilde M_{s,0}^2(k_i) \, \Theta\left( v_i < \eps v
  \right)} = &  \sum_{N=0}^\infty \frac{1}{N!}\prod_{i=1}^N [\df k_i] \tilde M_{s,0}^2(k_i) \, \Theta( v_i > \eps v) 
\,.
\end{align}
This gives the final equation
\begin{align}
\label{eq:fullF_NLL_final}
\Df\w_s\,  \fullF^{'\rm NLL}_s( \w_s, v)  =  & \Delta_s^{\rm LL}( v, \eps v)
 \sum_{N=0}^\infty \frac{1}{N!}\prod_{i=1}^N [\df k_i] \tilde M_{s,0}^2(k_i) \, \Theta( v_i > \eps v) 
\,.
\end{align}

In Sec.~\ref{sec:transfer_fully_differential_explicit_NLL} we will use
the SCET expression for the matrix element $\tilde M_{s,0}^2(k)$ and a
particularly suitable parametrization of the phase space integration
to derive an explicit expression for this result that can be
implemented in a straightforward numerical algorithm.

\subsection{General expression for the transfer function at NNLL}
\label{sec:transfer_fully_differential_general_NNLL}

The NNLL transfer function necessitates the calculation of
$\delta \sigma_F / \delta \w_F$ and
$\Sigma_F^{\rm max}( v)$ to NLL accuracy. This requires higher
order corrections to the correlated clusters used for the soft
transfer function, as well as the tree level single-emission amplitude
for the collinear transfer function. We will discuss these two cases
in turn.
\subsubsection{The soft transfer function}\mbox{}\\
\label{sec:transfer_fully_differential_general_NNLL_soft}
As discussed in Section~\ref{sec:logcounting}, in order to compute the
cumulative distribution beyond NLL, it is sufficient to expand the
NNLL correction perturbatively for fixed $\alpha_s \ln v$. In the
context of the soft transfer function, this amounts to correcting the NLL
result by adding either a single one-loop correction to the 1-particle
correlated cluster, or adding a single 2-particle correlated cluster
at tree level. This allows one to write the numerator of the fully
differential transfer function as
\begin{align}
\label{eq:diffSigma_NLL_1}
\frac{\delta\sigma_s^{\rm NLL}}{ \delta\w_s}   &= \frac{\delta\sigma_s^{\rm LL}}{ \delta\w_s} \left[ 1 + {\cal V}_s^{(1)}\right] + {\cal V}_s^{\rm LL}
\sum_{N=1}^\infty 
\sum_a \prod_{\substack{i=0 \\ i\neq a}}^N \tilde M_{s,0}^2(k_i)
\nn
& \qquad \times 
\Bigg[\tilde M_{s,1}^2(k_a)
+ \sum_{b>a} \tilde M_{s,0}^2(k_{a}, k_b)  \Bigg]
\,,
\end{align}
where we have expanded the virtual contribution beyond LL accuracy
perturbatively
${\cal V}_s^{\rm NLL} / {\cal V}_s^{\rm LL} = 1 + {\cal V}_s^{(1)} +
\ldots$
just as for the real emission matrix elements. As for the LL case, the
precise form of the virtuals is not important, as these will drop out
of the final result.

In \eq{diffSigma_NLL_1} there are IRC divergences in all of the three
terms of the r.h.s., and one needs to work out their
cancellation. 
We start by considering the denominator $\Sigma_{\max}$ in the
definition of the transfer function of eq.~\eqref{eq:transfer_NkLL_2},
and we will return to the general cancellation (which is required for
the numerator of the transfer function)
later. Using~\eq{Sigma_max_F_def} one finds
\begin{align}
\label{eq:diffSigma_max_NLL_2}
&\Sigma^{\rm max, NLL}_{s}( v) 
= \Sigma^{\rm max, LL}_{s}( v) \Bigg[ 1 + {\cal V}_s^{(1)} +  \int[\df k_a]
\tilde M_{s,1}^2(k_a) \Theta(\vtS(k_a) < v) 
\nn
&\qquad\qquad+ 
\int[\df k_a][\df k_b] \, S(2)\tilde M_{s,0}^2(k_a, k_b) \Theta(\vtS(k_a + k_b) < v)  \Bigg]
\,.
\end{align}
The main steps to arrive at this result involved handling the symmetry
factors in the phase space integration. First, we have used that the
number of permutations possible in summing over $a$ in the term
proportional to $\tilde M_{s,1}^2(k_a)$, and $b>a$ in the term
proportional to $\tilde M_{s,0}^2(k_a, k_b)$ of \eq{diffSigma_NLL_1}
are such that they convert the symmetry factor $S(N)$ arising from the
integration over the $N$-body phase space (see \eq{Dwn}) into $S(N-1)$
and $S(N-2) \, S(2)$, respectively. This can easily be seen in the
case where all final state particles are gluons, in which case
$S(N) = 1 / N!$ and the number of permutation is ${N \choose 1} = N$
and ${N \choose 2} = N(N-1) / 2 = N(N-1) S(2)$, respectively.
Finally, we have performed the shift $N \to M+1$ and $N \to M+2$ and
summed over $M$ to obtain the definition of $\Sigma^{\rm max, LL}_s$
given in \eq{SigmaMaxLL_def}.

In the square bracket of \eq{diffSigma_max_NLL_2}, the IRC divergences cancel between the various contributions. 
We start by canceling the divergence between the double real and real-virtual terms. For this, we use the well known result
\begin{align}
\label{eq:CMW_def}
&  M_{s,0}^2(k)  + \tilde M_{s,1}^2(k)+ \int [\df k_a][\df k_b] \delta^{(3)}(k - k_{a} - k_{b}) S(2)\tilde M_{s,0}^2(k_a, k_b) 
\nn
& \qquad = M_{s,0}^2(k) \left(1 + K \, \frac{\alpha_s(k_\perp)}{2\pi}+\dots\right)\,,
\end{align}
where the $\dots$ represent terms beyond NNLL, and $K$ is given by the
ratio of the 2-loop to the 1-loop cusp anomalous dimension
\begin{align}
K & = \frac{\Gamma^{(2)}}{\Gamma^{(1)}} = C_A \left( \frac{67}{18} - \frac{\pi^2}{6} \right) - \frac{5}{9} n_f
\,.
\end{align}
To obtain \eq{CMW_def} it is crucial that the coupling in
$M_{s,0}^2(k)$ is evaluated at $\mu = k_\perp$, which, as already
discussed, is the only possible choice of scales in SCET for this
quantity. While this result is typically derived in a setting outside
of SCET~\cite{Catani:1990rr,Banfi:2018mcq,Catani:2019rvy}, we have
explicitly checked that it is obtained using the Feynman rules of the
SCET soft Lagrangian.

Combining \eqs{diffSigma_max_NLL_2}{CMW_def}, we obtain
\begin{align}
\label{eq:diffSigma_max_NLL_3}
\Sigma^{\max, {\rm NLL}}_{s}( v) = \Sigma^{\max, {\rm LL}}_{ s}( v) \left[ 1 + {\cal V}_s^{(1)} + \int\![\df k]\,\tilde M_{s,0}^2(k) K\,\frac{\alpha_s(k_\perp)}{2\pi}\Theta(\vtS(k) < v) \right]
\,.
\end{align}
This gives the denominator of the general expression for the transfer
function of~\eq{transfer_NkLL_2}, as well as the Sudakov factor at NLL accuracy
\begin{align}
\label{eq:Delta_NLL}
\Delta_s^{\rm NLL}( v , \eps v) = \Delta_s^{\rm LL}( v, \eps v) \left[1 - \int\![\df k] \, K \, \tilde M_{s,0}^2(k) \frac{\alpha_s(k_\perp)}{2\pi}\Theta(\eps v < \vtS(k) < v) \right]
\,.
\end{align}

The only ingredient missing from \eq{transfer_NkLL_2} is the numerator of the second term. Using again \eq{diffSigma_NLL_1} we can write
\begin{align}
\label{eq:NNLL_numerator}
\Df\w_s \frac{\delta\sigma_s^{\rm NLL}}{ \delta\w_s}
&=  {\cal V}_s^{\rm LL}\sum_{N=0}^\infty S(N)\prod_{i=1}^N [\df k_i] \tilde M_{s,0}^2(k_i)
\Bigg[ 1 + {\cal V}_s^{(1)}
\nn
&\qquad+  [\df k_a]
M_{s,1}^2(k_a) 
+ 
[\df k_a][\df k_b] \, S(2)\tilde M_{s,0}^2(k_a, k_b)  \Bigg]
\,.
\end{align}
Note that the first, second and third terms in the square bracket contain $N$, $N+1$ and $N+2$ momenta, respectively, and the action of the observable therefore is different in each of these terms, which are again separately IRC divergent.

The IRC divergence of ${\cal V}_s^{(1)}$ ultimately cancels in the ratio to $\Sigma_{\max}$ in \eq{diffSigma_NLL_1}, so we ignore it for the time being. To handle the cancellation of the IRC divergence between the second and third term, we introduce a simple subtraction term
\begin{align}
[\df k_a][\df k_b] \, S(2) \, \tilde M_{s,0}^2(k_a, k_b) \, {\cal P}_+(k_a, k_b)
\end{align}
in the square bracket, where the projection ${\cal P}_+(k_a, k_b)$ is defined through its action on the observable $V(q_n, q_\bn; \{k_i\}, k_a, k_b)$ upon integration over the phase space of $k_a$ and $k_b$ as
\begin{align}
&\int [\df k_a][\df k_b] \, S(2) \, \tilde M_{s,0}^2(k_a, k_b) {\cal P}_+(k_a, k_b)  \Theta(V(q_n, q_\bn;\{k_i\}, k_a, k_b) < v)
\\
& \qquad \equiv \int \! [\df k] [\df k_a][\df k_b] \, S(2)\, \tilde M_{s,0}^2(k_a, k_b) \,  \delta^{(3)}(k - k_{a} - k_{ b}) \, \Theta(V(q_n, q_\bn;\{k_i\}, k) < v)
\nn
& \qquad \equiv \int \! [\df k_a][\df k_b] \, S(2)\, \tilde M_{s,0}^2(k_a, k_b) \,   \Theta(V(q_n, q_\bn;\{k_i\},  k_{ab})  < v)
\notag
\,,
\end{align}
where $\{k_i\}$ represent a set of other emissions, and $k_{ab}$ is
the massless momentum that is constructed from the transverse momentum
and pseudo-rapidity of the vector $k_a+k_b$.

With this subtraction term, \eq{NNLL_numerator} can be recast as
\begin{align}
\label{eq:NNLL_numerator_2}
\Df\w_s \frac{\delta\sigma_s^{\rm NLL}}{ \delta\w_s}
&=  {\cal V}_s^{\rm LL}\sum_{N=0}^\infty S(N)\prod_{i=1}^N [\df k_i] \tilde M_{s,0}^2(k_i) 
\nn
&\qquad\times
\Bigg[ 1 + {\cal V}_s^{(1)}
+  [\df k_{a}] \, K \,\tilde M_{s,0}^2(k_{a}) \frac{\alpha_s(k_{a,\perp})}{2\pi} 
\nn
&\qquad \qquad+   
[\df k_a][\df k_b] \, S(2) \tilde M_{s,0}^2(k_a, k_b) \left[ 1 - {\cal P}_+(k_a, k_b) \right] \Bigg]\,,
\end{align}
where we have used \eq{CMW_def} to combine the integrated subtraction with $\tilde M_{s,1}^2(k_a)$. 

Combining \eqs{diffSigma_max_NLL_3}{NNLL_numerator_2}, the virtual
correction drops out again, and one obtains after a few lines of
algebra the following soft transfer function
\begin{align}
\label{eq:fullFNNLL_2}
\Df\w_s \fullF_s^{'\rm NNLL}( \omega_s) 
&
=  \Delta_s^{\rm NLL}( v, \eps v) 
\sum_{N=0}^\infty \frac{1}{N!}\prod_{i=1}^N [\df k_i] \, \tilde M_{s,0}^2(k_i) \, \Theta(v_i > \eps v)
\nn
&\qquad\ \times \Big[1 
+ [\df k_a] \, K \,  \tilde M_{s,0}^2(k_a) \, \frac{\alpha_s(k_{a,\perp})}{2\pi} \, \Theta(v_a > \eps v)
\nn
&\qquad\qquad
+   
[\df k_a][\df k_b] \, \tilde M_{s,0}^2(k_a,k_b) \left[ 1 - {\cal P}_+(k_a, k_b) \right]\Bigg]
\,,
\end{align}
and we have taken the limit $\eps \to 0$ in the term that is now regulated through the ${\cal P}_+$ projection. 
As a final step, we replace the $\Delta_s^{\rm NLL}$ with the expression given in \eq{Delta_NLL}. The final result can be conveniently expressed as
\begin{align}
\label{eq:FNNLLSoft2}
\Df\w_s \fullF_s^{'\rm NNLL}( \omega_s) &= \Df\w_s \fullF_s^{'\rm NLL}( \omega_s) + \Df\w_s \delta \fullF_s^{'\rm NNLL}( \omega_s)
\nn
&
=  \Delta_s^{\rm LL}( v, \eps v) 
\sum_{N=0}^\infty \frac{1}{N!}\prod_{i=1}^N [\df k_i] \, \tilde M_{s,0}^2(k_i) \, \Theta(v_i > \eps v)
\nn
&\qquad \times \Bigg[1 
+ [\df k_a] \, K\,  \tilde M_{s,0}^2(k_a) \, \frac{\alpha_s(k_{a,\perp})}{2\pi} \, \left[ 1 - {\cal P}_0(k_a) \right]
\nn
&\qquad\qquad
+   
[\df k_a][\df k_b] \, \tilde M_{s,0}^2(k_a,k_b) \left[ 1 - {\cal P}_+(k_a, k_b) \right] \Bigg]
\,.
\end{align}
Here we have defined a second projection operator again through its action on the observable $V[q_n, q_\bn;\{k_i\}, k_a]$ upon integration over the phase space of $k_a$ as
\begin{align}
\label{eq:P0_def}
&\int [\df k_a] \, \tilde M_{s,0}^2( k_a) {\cal P}_0(k_a)  \Theta(V(q_n, q_\bn;\{k_i\}, k_a) < v)
\nn
& \qquad
\equiv \int \! [\df k_a]\, \tilde M_{s,0}^2(k_a) \,   \Theta(V(q_n, q_\bn;\{k_i\}) < v) \, \Theta(V_{\max}(k_a) < v)
\,,
\end{align}
and, as above, we have taken the limit $\eps \to 0$ in the term regulated by the ${\cal P}_0$ projection. 

This is the final result for the fully differential soft transfer function at NNLL written in terms of a general expression for the correlated clusters. In Section \ref{sec:transfer_fully_differential_explicit_NLL} we will use the explicit SCET results together with a suitable phase space parametrization to simplify these results further. 

\subsubsection{The collinear transfer function}\mbox{}\\
\label{sec:transfer_fully_differential_general_NNLL_collinear}
Since, as already discussed, collinear emissions only start contributing at NNLL order to the transfer function, it is sufficient to include the effect of one single emission at this order. One can therefore write
\begin{align}
\Df\w_n \, \frac{\delta\sigma_n^{\rm NLL}}{ \delta\w_n} = \left[ 1 +
  {\cal V}_n^{(1)}\right] \delta(\w_n - \w[q_{n}]) + [\df k_n] \, \tilde M_{n,0}^2(k_n) \,,
\end{align}
where the subscript $n$ denotes the direction of the collinear sector. The $\Sigma^{\max}_ {n}$ is given by
\begin{align}
\Sigma^{\max}_ {n}(v) &= 1 + {\cal V}_n^{(1)}+ \int\! [\df k_n]  \, \tilde M_{n,0}^2(k_n) \, \Theta( V_{\max}(k_n) < v)
\,.\end{align}
Using \eq{transfer_NkLL_2}, and following the same steps as in the soft case, one finds
\begin{align}
\label{eq:fullFNNLL_coll}
\Df \w_n \fullF^{'\rm NNLL}_n( \w_n) &= \Df\w_n \fullF_n^{'\rm NLL}(\omega_n) + \Df\w_n \delta \fullF_n^{'\rm NNLL}( \omega_n)
\nn
&= 1 +  [\df k_n]  \, \tilde M_{n,0}^2(k_n) \left[ 1 - {\cal P}_0(k_n) \right]
\,,
\end{align}
where ${\cal P}_0(k_n)$ was defined in \eq{P0_def}. We stress that the
collinear sector must undergo a zero-bin subtraction to avoid double
counting with the soft sector. This will be directly performed in
Section~\ref{sec:transfer_fully_differential_explicit_NNLL_collinear},
where the explicit parametrisation of the collinear transfer function
is given.

\section{Explicit results for transfer function}
\label{sec:explicit_transfer}

In this section we formulate the somewhat abstract results obtained in
the previous section in a concrete way that is suitable for a
numerical evaluation.\footnote{The resulting numerical algorithms are
  presented in Appendix~\ref{app:Numerical_algorithms}.}~We rely on
the explicit parametrization of the phase space and SCET matrix
elements reported in Appendices~\ref{app:Phase_space_parametrization}
and~\ref{app:SCET_squared_amplitudes}.

We start by making some general remarks. The considerations of the
previous section have resulted in equations where we have neglected
squared amplitudes which are of subleading logarithmic order. However,
each of the transfer functions given in
Eqs.~\eqref{eq:fullF_NLL_final}, ~\eqref{eq:FNNLLSoft2}
and~\eqref{eq:fullFNNLL_coll}, as well as their combination into the
final transfer function according to \eq{transfer-nonfact}, still
contain subleading logarithmic terms. These essentially originate from
the following three sources
\begin{enumerate}
\item Higher logarithmic contributions to the running coupling constant
\item Details of the phase space boundaries
\item Details of the observable used in the final combination
\end{enumerate}
The first source of subleading corrections is due to the presence of
the coupling constant in each correlated cluster. One obviously needs
to evaluate the running coupling at the desired resummation accuracy,
i.e. only include $\beta_0$ at NLL, add $\beta_1$ at NNLL and so
on. The second source is due to the rapidity boundaries, which in
general are such that each emission develops a subleading logarithmic
contribution upon phase space integration. The third source is arising
from the fact that the transfer functions are integrated against the
observable's measurement function. In general, this integral will
again give rise to subleading logarithmic and power suppressed
effects.

As we will show in the following, it is possible to address each of
the above points in such a way that no subleading logarithms are
contained in the final transfer function. 
While this sounds very different from what is done in the usual SCET
resummation, it is in fact quite similar. An SCET resummation relies
on fixed order computations of anomalous dimensions (extracted from
the divergent structure of the result) as well as finite
contributions. In these calculations, regulators and observable
constraints are expanded around their limiting value. To a given order
of resummation, one deliberately picks the anomalous dimensions and
finite pieces at different orders from the fixed order calculation. In
a numerical procedure as employed here, one then needs to find a way
to reproduce these results and to ensure that no subleading logarithms
are produced upon phase space integration.

\subsection{Explicit expressions at NLL}
\label{sec:transfer_fully_differential_explicit_NLL}

The final expression of the transfer function in
\eq{transfer_NLL_nonfact} is given in terms of \eq{fullF_NLL_final},
with the Sudakov defined in \eq{DeltaLL_def}. One therefore needs to
parametrise the phase space integral over the single particle soft
correlated cluster. Using the expressions reported in
Appendices~\ref{app:Phase_space_parametrization}
and~\ref{app:SCET_squared_amplitudes}, we can write
\begin{align}
  \label{eq:NLL_integrand_1}
  [\df k_i] \, \tilde M_{s,0}^2(k_i) = \sum_{\ell} \frac{\df v_i}{v_i}  \df \chi_i \, \frac{\df \phi_i}{2\pi} \,\frac{2C_\ell}{a} \, \frac{\alpha_s(k_{i,\perp})}{\pi} \, \frac{\ln 1/v_\ell}{a+b_\ell} 
  \,,\end{align}
where $C_\ell$ is the colour factor associated with leg $\ell$.
Note that in this expression it should be understood that the
variables $v_i$, $\chi_i$ and $\phi_i$ are always defined with respect
to the leg $\ell$ relative to the emission $k_i$, and we only suppress this dependence here to simplify the notation. 
To NLL accuracy, \eq{NLL_integrand_1} can be simplified by performing the following Taylor expansion
\begin{align}
\label{eq:splittingExpansion}
\frac{2C_\ell}{a} \, \frac{\alpha_s(k_{i,\perp})}{\pi} \, \frac{\ln 1/v_\ell}{a+b_\ell}  & = P^{\rm LL}_{s,\ell}(v, \chi_i) + \frac{\df P^{\rm LL}_{s,\ell}(v, \chi_i)}{\df \ln 1/v} \ln \frac{v}{v_i}
\nn
& \qquad  - P^{\rm LL}_{s,\ell}(v,\chi_{i}) \frac{\alpha_s^{\rm LL}(\kappa_{i,\perp})}{4\pi} \frac{\beta_1}{\beta_0} \ln (1+t) + \ldots
\,,\end{align}
where
\begin{align}
\label{eq:Pl_exp}
P^{\rm LL}_{s,\ell}(v, \chi_i) = \frac{2C_\ell}{a} \, \frac{\ln 1/v}{a+b_\ell} \, \frac{\alpha_s^{\rm LL}(\kappa_{i,\perp})}{\pi}\,, \qquad \kappa_\perp = Q \, e^{\frac{\ln 1/v}{a} \left(\frac{b}{a+b}\,\chi_i - 1\right)}
\,,
\end{align}
and the running coupling at LL accuracy is simply given by
\begin{align}
\label{eq:alphas_def}
\alpha_s^{\rm LL}(\kappa_{i,\perp}) \equiv \frac{\alpha_s(Q)}{1+t}\,, \qquad t = \frac{\alpha_s(Q)}{4\pi} \beta_0 \ln \frac{\kappa_{i,\perp}^2}{Q^2}
\,.
\end{align}
One can easily see that the integration over the second term leads to
a subleading logarithmic contribution, which can be dropped to the
order we are working. It will be included in the next section when
going to NNLL.

Using this, one immediately obtains for the Sudakov factor in \eq{DeltaLL_def}
\begin{align}
\label{eq:Delta_LL_1}
\Delta_s^{\rm LL}( v, \eps v) 
&= \exp \left\{ - \int_{\eps v}^v  \! \frac{\df \bar v}{\bar v} \int_{0}^{1} \! \df \chi \,  \int_{0}^{2\pi} \! \frac{\df \phi}{2\pi} \, \sum_{\ell}P^{\rm LL}_{s,\ell}(v; \chi)  \right\}
\,,
\end{align}
that can be computed analytically yielding
\begin{align}
\label{eq:Delta_LL_3}
\Delta_s^{\rm LL}( v, \eps v)  &= e^{\sum_\ell P^{\rm LL}_{s,\ell}(v)\ln \delta } = \delta^{\sum_\ell P^{\rm LL}_{s,\ell}(v)} 
\,,
\end{align}
with
\begin{align}
P^{\rm LL}_{s,\ell}(v) & \equiv \int_{0}^{1} \! \df \chi' \, P^{\rm LL}_{s,\ell}(v; \chi')= 4 C_\ell \, \frac{\ln[1-\frac{2\lambda}{a+b}] - \ln[1-\frac{2\lambda}{a}]}{b \, \beta_0} \,, \quad \lambda =  \frac{\alpha_s(Q)}{4\pi} \beta_0 \ln 1/v
\,.
\end{align}
Combining these results into \eq{transfer_NLL_nonfact}, one finds
\begin{align}
\label{eq:NLL_transfer_final}
\fullF^{\rm NLL}(\Phi_B; v)  = &\Delta_s^{\rm LL}( v, \eps v)
                              \,\left[ 1 + \sum_{N=1}^\infty \frac{1}{N!}\prod_{i=1}^N  \sum_{\ell} \int_{\eps v} \frac{\df v_i}{v_i} \, \int \!  \df \chi_i  \,P^{\rm LL}_{s,\ell}(v; \chi_i) \, \int \! \frac{\df \phi_i}{2\pi} \right]\notag\\
&\times 
\Theta(V(q_n, q_\bn; k_1, \ldots, k_N) < v)\notag\\
&\equiv \fullF^{\rm NLL}\left[P^{\rm LL}_{s,\ell}, V\right](\Phi_B; v) 
\,.
\end{align}
In the last line we have made explicit that the functional form of
$\fullF^{\rm NLL}(\Phi_B; v)$ is determined entirely in terms of
$P^{\rm LL}_{s,\ell}$ and the observable $V$.

\subsection{Explicit expressions at NNLL}
\label{sec:transfer_fully_differential_explicit_NNLL}

In this section we derive explicit expressions for the NNLL
corrections to the transfer function. They can be classified in 3
broad terms. First, one needs to re-compute the NLL transfer function
including the higher order terms to the various expansions that were
made in~\eq{splittingExpansion}. One therefore writes
\begin{align}
\label{eq:fullF_NNLL_NLL}
\fullF^{\rm NLL}\left[P^{\rm LL}_{s,\ell}, V\right](\Phi_B; v) \to \fullF^{\rm NLL}\left[P^{\rm NLL}_{s,\ell}, V\right](\Phi_B; v)
\,.
\end{align}
As previously discussed, to obtain NNLL accuracy it is sufficient to
describe a single emission with the NLL splitting function, while
using LL splitting functions for all other emissions. Therefore, one
could expand $P^{\rm NLL}_{s,\ell}$ about $P^{\rm LL}_{s,\ell}$ in
Eq.~\eqref{eq:fullF_NNLL_NLL}, and retain only the linear term at this
accuracy, i.e.
\begin{align}
\label{eq:fullF_NNLL_NLL_expanded}
\fullF^{\rm NLL}\left[P^{\rm NLL}_{s,\ell}, V\right](\Phi_B; v) =
  \fullF^{\rm NLL}\left[P^{\rm LL}_{s,\ell}, V\right](\Phi_B; v) +
  \delta\fullF^{\rm NLL}\left[P^{\rm LL}_{s,\ell}, P^{\rm NLL}_{s,\ell},
  V\right](\Phi_B; v) + \ldots
\,,
\end{align}
where the dots represent subleading logarithmic corrections. We leave
the details of this operation to Appendix~\ref{app:expansion_PNLL},
and only present the expression for $P^{\rm NLL}_{s,\ell}$ below. We
stress that this approximation is by no means mandatory, and one can
directly evaluate the NLL transfer function with the NLL splitting
function.\footnote{The corresponding Monte Carlo algorithm is given in
Appendix~\ref{sec:NNLL_corrections_to_FNLL_algorithms}.}

As a second NNLL term, we need the higher order corrections to the
soft n-particle correlated clusters, given by in \eq{FNNLLSoft2}. As
discussed there, this involves either one or two soft emissions on top
of the ones required for the NLL transfer function. Finally, one needs
the collinear transfer function, given in \eq{fullFNNLL_coll}. As
before, we will denote these two corrections by
\begin{align}
\delta \fullF^{\rm NNLL}_{F}[V](\Phi_B; v)
\,,
\end{align}
with $F \in \{s, n,\bn\}$. Here we explicitly state the dependence on
the observable, as this will become important later.  Since one only
needs to include one correction at a time, the NNLL corrections to the
soft transfer function are always combined with the NLL result for the
collinear transfer function (which is just a delta function), while
the NNLL correction to the collinear transfer function is combined
with the NLL soft transfer function derived the previous section. We
now discuss these three pieces in turn.

\subsubsection{Expression for $P^{\rm NLL}_{s,\ell}$}
In this subsection we just provide the expression for the NLL
splitting function $P^{\rm NLL}_{s,\ell}(v, \chi_a)$, described
above. This is simply defined by keeping the NLL term in the expansion
\eq{splittingExpansion}, that is
\begin{align}
\label{eq:NLL_splitting}
& P^{\rm NLL}_{s,\ell}(v, v_a, \chi_a, \phi_a) 
\\
& \qquad = P^{\rm LL}_{s,\ell}(v, \chi_a) 
+ \left[ \frac{\df P^{\rm LL}_{s,\ell}(v,\chi_{a})}{\df \ln 1/v} \ln \frac{v}{v_{a}} - \frac{\beta_1}{\beta_0} \, P^{\rm LL}_{s,\ell}(v,\chi_{a}) \frac{\alpha_s^{\rm LL}(\kappa_\perp)}{4\pi}  \ln (1+t) \right] 
\notag
\,.
\end{align}

To use this result in \eq{NLL_transfer_final}, we need the
integral of  \eq{NLL_splitting} over $\chi_a$ and $\phi_a$. One finds
\begin{align}
P^{\rm NLL}_{s,\ell}(v, v_a) &=  P^{\rm LL}_{s,\ell}(v) + \frac{\df P^{\rm
  LL}_{s,\ell}(v)}{\df \ln 1/v}  \ln \frac{v}{v_a} 
 +P_{s,\ell}^{\rm NLL, \beta}(v)\notag\\
&\equiv P^{\rm LL}_{s,\ell}(v) + \frac{\df P^{\rm
  LL}_{s,\ell}(v)}{\df \ln 1/v}  \ln \frac{v}{v_a}
 +P_{s,\ell}^{\rm NLL, \beta}(v)
\,,
\end{align}
where we defined
\begin{align}
\frac{\df P^{\rm LL}_{s,\ell}(v)}{\df \ln 1/v} &= \frac{\alpha_s(Q)}{\pi} \, \frac{2 C_\ell}{(a-2\lambda)(a+b-2\lambda)} 
\nn
P_{s,\ell}^{\rm NLL, \beta}(v) &= -\frac{\beta_1}{\beta_0}
                             \frac{\alpha_s(Q)}{\pi} \, C_\ell \,
                             \frac{a(a+b-2\lambda)\ln
                             \left[1-\frac{2\lambda}{a} \right] -
                             (a+b)(a-2\lambda)\ln
                             \left[1-\frac{2\lambda}{a+b} \right] + 2
                             b \lambda}{b
                             \beta_0(a-2\lambda)(a+b-2\lambda)}
\,.
\end{align}

\subsubsection{$\delta \fullF_{s}^{'\rm NNLL}( v)$: The soft contribution}
\label{sec:transfer_fully_differential_explicit_NNLL_soft}
We recall once more that in order to obtain a NNLL result for the
cumulative distribution, the NNLL correction to the soft transfer
function is to be combined with the NLL collinear transfer function,
which at this order is just a delta function. Starting from
\eq{FNNLLSoft2} and keeping only the term involving a single extra
emission we find, using eq.~\eqref{eq:splittingExpansion}
\begin{align}
\frac{2C_\ell}{a} \, \frac{\alpha^2_s(k_{i,\perp})}{2\pi^2} \, \frac{\ln 1/v_\ell}{a+b_\ell} \, K  = P^{\rm LL}_{s,\ell}(v; \chi_a) \, K \, \frac{\alpha_s^{\rm LL}(\kappa_{a,\perp})}{2\pi} + \ldots
\end{align}
Here we have performed the same expansion as in the NLL case, and
$\ldots$ denote terms that start to contribute at N${}^3$LL. Plugging
this back into eq.~\eqref{eq:FNNLLSoft2} gives for the correction with
a single emission
\begin{align}
\label{eq:NNLL_transfer_soft_1}
& \delta \fullF^{\rm NNLL}_{s,1}[V](\Phi_B; v) = \Delta_s^{\rm LL}( v)  \left[ 1 + \sum_{N=1}^\infty \frac{1}{N!}\prod_{i=1}^N  \sum_{\ell} \int_{\eps v} \frac{\df v_i}{v_i} \, \int \!  \df \chi_i  \,P^{\rm LL}_{s,\ell}(v; \chi_i) \, \int \! \frac{\df \phi_i}{2\pi} \right]
\nn
& \qquad \times \sum_\ell \int \! \frac{\df v_{a}}{v_a} \int  \!\df \chi_{a} \int  \!\frac{\df \phi_{a}}{2\pi} \, P^{\rm LL}_{s,\ell}(v; \chi_a) \, K \, \frac{\alpha_s^{\rm LL}(\kappa_{a,\perp})}{2\pi}
\nn
& \qquad \times
\Bigg[ \Theta(V(q_n, q_\bn; k_1, \ldots, k_N, k_a) < v) -  \Theta(V(q_n, q_\bn; k_1, \ldots, k_N) < v) \,  \Theta(V_{\max}(k_a) < v)\Bigg]\,
\,.
\end{align}

For the term involving two extra emissions, we use the phase space parametrization given in Appendix \ref{app:Phase_space_parametrization} and the soft 2-particle correlated cluster given in Appendix \ref{app:SCET_squared_amplitudes}. 
We can write
\begin{align}
\label{eq:Mab_def}
&[\df k_a][\df k_b] \tilde M_{s,0}^2 (k_a, k_b) 
\nn
&\quad = \sum_{\ell_a  = \ell_b} \frac{1}{4\pi^2} \frac{1}{a}
  \frac{\df v_{\ell_a}}{v_{\ell_a}} \frac{\df
  \kappa}{\kappa} \, \df \Delta\eta \,\df \chi_{\ell_a} \, \frac{\df
  \phi_{\ell_a}}{2\pi} \,\frac{\df \phi_{\ell_b}}{2\pi} \,
  k_{\perp, a}^4 \, \kappa^2 \,\eta_{\rm max}\notag\\
&\qquad\times\alpha_s(k_{\perp,a})\,
  \alpha_s(k_{\perp,b}) {\hat M}^2_{s,0}(k_a,k_b)
\,,\end{align}
where $\hat{M}_{s,0}^2 (k_a, k_b)$ is given in \eq{Mhat_SO}, and we defined
$\kappa = k_{\perp,b}/k_{\perp,a}$. As done for the
single emission contribution above, we now Taylor expand the arguments of the
couplings and the rapidity bounds in order to retain only NNLL terms,
hence obtaining
\begin{align}
\label{eq:weight_correlated}
&[\df k_a][\df k_b] \tilde M_{s,0}^2 (k_a, k_b) 
\nn
&\quad = \sum_{\ell_a  = \ell_b} \frac{1}{4\pi^2} \frac{1}{a}
  \frac{\df v_{\ell_a}}{v_{\ell_a}} \frac{\df
  \kappa}{\kappa} \, \df \Delta\eta \,\df \chi_{\ell_a} \, \frac{\df
  \phi_{\ell_a}}{2\pi} \,\frac{\df \phi_{\ell_b}}{2\pi} \,
  k_{\perp, a}^4 \, \kappa^2 \frac{1}{a+b_{\ell_a}} \ln \frac{1}{v}\notag\\
&\qquad\times\left(\alpha_s^{\rm LL}(\kappa_{a,\perp})\right)^2{\hat M}^2_{s,0}(k_a,k_b)
+ \ldots
\,,
\end{align}
where $\ldots$ refers to higher logarithmic terms. Using
eq.~\eqref{eq:FNNLLSoft2} leads to
\begin{align}
\label{eq:NNLL_transfer_soft_2}
& \delta \fullF^{\rm NNLL}_{s,2}[V](\Phi_B; v) =  \Delta_s^{\rm LL}( v)  \left[ 1 + \sum_{N=1}^\infty \frac{1}{N!}\prod_{i=1}^N  \sum_{\ell} \int_{\eps v} \frac{\df v_i}{v_i} \, \int \!  \df \chi_i  \,P^{\rm LL}_{s,\ell}(v; \chi_i) \, \int \! \frac{\df \phi_i}{2\pi} \right]
\nn
& \qquad \times \int \! [\df k_a][\df k_b] \tilde M_{s,0}^2 (k_a, k_b)
\nn
& \qquad \times
\Bigg[ \Theta(V(q_n, q_\bn; k_1, \ldots, k_N, k_a, k_b) < v) -  \Theta(V(q_n, q_\bn; k_1, \ldots, k_N, k_{ab}) < v) \Bigg]
\,,
\end{align}
where the second line is as in \eq{Mab_def}, and as before $k_{ab}$ is
the massless momentum that is constructed from the transverse momentum
and pseudo-rapidity of the vector $k_a+k_b$.

The total NNLL result originating from the higher order soft correlated clusters is then
\begin{align}
\label{eq:fullFNNLL_S}
\delta \fullF^{\rm NNLL}_s[V](\Phi_B; v)  = \delta \fullF^{\rm NNLL}_{s,1}[V](\Phi_B; v)  + \delta \fullF^{\rm NNLL}_{s,2}[V](\Phi_B; v) 
\,.
\end{align}

\subsubsection{$\delta \fullF_{n,\bn}^{'\rm NNLL}( v)$: The
  collinear contribution }
\label{sec:transfer_fully_differential_explicit_NNLL_collinear}
The NNLL collinear correction is obtained by combining the fully differential NNLL collinear transfer function with the NLL soft transfer function, obtained in Section~\ref{sec:transfer_fully_differential_explicit_NLL}. Using the collinear phase space parametrization and collinear matrix elements given in Appendices~\ref{app:Phase_space_parametrization} and~\ref{app:SCET_squared_amplitudes}, respectively, we obtain 
\begin{align}
[\df k_n] \left|M_{n}^{(0)}(k_n)\right|^2 &=   \frac{\df v_n}{v_n} \, \frac{\df z_n}{z_n}  \, \frac{\df \phi_n}{2\pi} \frac{C_F}{a} \frac{\alpha_s(k_{n,\perp})}{\pi}  \, \left[(1-z_n)^2+1\right]\,.
\nn
&= \frac{\df v_n}{v_n} \,\df z_n  \, \frac{\df \phi_n}{2\pi} \, P^{\rm NLL}_{n}(v, z_n) + \ldots
\,,
\end{align}
where
\begin{align}
\label{eq:hard_collinear_splitting_function}
P^{\rm NLL}_{n}(v, z_n) \equiv  \frac{C_F}{a+b_\ell} \, \frac{\alpha_s^{\rm LL}(Q\, v^{1/(a+b)})}{\pi} \,  \, \frac{(1-z_n)^2+1}{z_n} 
\,.
\end{align}
As for the soft transfer function, we have expanded the integrand
retaining only the most singular terms. 
Thus, we find
\begin{align}
\Df \w_n \fullF^{'\rm NNLL}_n( \w_n) & = 1 +  \frac{\df v_n}{v_n} \, \df z_n  \, \frac{\df \phi_n}{2\pi} \, P^{\rm NLL}_{n}(v, z_n) \, \left[ 1 - {\cal P}_0(k_n) \right] 
\,.
\end{align}

To obtain the final collinear contribution to the NNLL transfer
function, we first need to combine this with the fully differential
soft transfer function at NLL obtained in
Section~\ref{sec:transfer_fully_differential_explicit_NLL}. This
gives, for each collinear sector,
\begin{align}
\label{eq:NNLL_transfer_coll}
& \delta \fullF^{\rm NNLL}_{n}[V](\Phi_B; v) = \Delta_s^{\rm LL}( v)  \left[ 1 + \sum_{N=1}^\infty \frac{1}{N!}\prod_{i=1}^N \sum_{\ell} \int_{\eps v} \frac{\df v_i}{v_i} \, \int \!  \df \chi_i  \,P^{\rm LL}_{s,\ell}(v; \chi_i) \, \int \! \frac{\df \phi_i}{2\pi} \right]
\nn
& \qquad \times \int \! \frac{\df v_a}{v_a} \int  \! \df z_a  \int  \!\frac{\df \phi_a}{2\pi} \, P^{\rm NLL}_{n}(v, z_a) 
\nn
& \qquad \times
\Bigg[ \Theta(V(q_n, q_\bn; k_1, \ldots, k_N, k_a) < v) -  \Theta(V(q_n, q_\bn; k_1, \ldots, k_N) < v) \,  \Theta(V_{\max}(k_a) < v)\Bigg]\,
\,.
\end{align}
Finally, we need to perform the 0-bin subtraction to avoid the double
counting with the soft sector. This amounts to subtracting from
eq.~\eqref{eq:NNLL_transfer_coll} its purely soft approximation,
defined as
\begin{align}
\label{eq:NNLL_transfer_0bin}
& \delta \fullF^{\rm NNLL}_{n,\, {\rm 0-bin}}[V](\Phi_B; v) = \Delta_s^{\rm LL}( v)  \left[ 1 + \sum_{N=1}^\infty \frac{1}{N!}\prod_{i=1}^N  \sum_{\ell} \int_{\eps v} \frac{\df v_i}{v_i} \, \int \!  \df \chi_i  \,P^{\rm LL}_{s,\ell}(v; \chi_i) \, \int \! \frac{\df \phi_i}{2\pi} \right]
\nn
& \qquad \times \int \! \frac{\df v_a}{v_a} \int  \! \df z_a  \int
  \!\frac{\df \phi_a}{2\pi} \, P^{\rm NLL}_{n,\,{\rm 0-bin}}(v, z_a) 
\nn
& \qquad \times
\Bigg[ \Theta(V(q_n, q_\bn; k_1, \ldots, k_N, k_a) < v) -
  \Theta(V(q_n, q_\bn; k_1, \ldots, k_N) < v) \,  \Theta(V_{\rm max}(k_a) < v)\Bigg]\,
\,,
\end{align}
where 
\begin{align}
\label{eq:0-bin_splitting_function}
P^{\rm NLL}_{n,\,{\rm 0-bin}}(v, z_n) \equiv  \frac{C_F}{a+b_\ell} \, \frac{\alpha_s^{\rm LL}(Q\, v^{1/(a+b)})}{\pi} \,  \, \frac{2}{z_n} 
\,,
\end{align}
and $V_{\rm max}(k_a)$ in eq.~\eqref{eq:NNLL_transfer_0bin} is to be
considered in the soft limit, i.e.
$V_{\rm max}(k_a)=\tilde{V}_s(k_a)$, defined in eq.~\eqref{eq:Vmax_0}.
The NNLL collinear correction is simply given by the difference of
eqs.~\eqref{eq:NNLL_transfer_coll} and~\eqref{eq:NNLL_transfer_0bin},
namely
\begin{equation}
\delta \fullF^{\rm NNLL}_{n, \,{\rm final}}[V](\Phi_B; v)  = \delta \fullF^{\rm NNLL}_{n}[V](\Phi_B; v) -\delta \fullF^{\rm NNLL}_{n,\, {\rm 0-bin}}[V](\Phi_B; v) \,.
\end{equation}

Note that the same definition of the observable is used in
both~\eqref{eq:NNLL_transfer_coll}
and~\eqref{eq:NNLL_transfer_0bin}. However, as we will discuss in the
next section, the whole observable in the zero-bin contribution has to
be expanded about its soft limit.
\section{Dealing with the observable}
\label{sec:observable}

The expressions for the transfer function derived in the previous
section use the generic rIRC safe observable
$V(q_n, q_\bn; k_1, \ldots, k_N)$. In this section we discuss how to
evaluate the observable on a set of emissions $k_i$ corresponding to a
given sector of the SCET Lagrangian.

One possible choice, convenient for a flexible MC implementation,
would be to use the full definition of the observable, which often is
know in the form of an algorithm. For example, a jet observable
requires a sequential clustering algorithm, and even a simple
observable such as thrust involves a minimization procedure to find
the thrust axis. An issue with using the full definition is that it
often relies on the fact that the final state satisfies momentum
conservation, which is explicitly invalidated when constructing the
kinematics of the momenta of the fully differential transfer
function. This is because phase space limits were explicitly taken in
the soft and collinear limits. Such a problem can be overcome by
reconstructing the final state kinematics after the generation of the
radiation, following a procedure similar in spirit to what is
currently used in some parton shower
generators~\cite{Gieseke:2003rz,Bahr:2008pv}. However, a drawback of
this approach is that the corresponding resummed prediction will
include uncontrolled subleading-logarithmic and subleading-power
corrections that are ultimately eliminated with the matching to fixed
order at the relevant perturbative accuracy.

A different recipe, commonly adopted in resummations, is to expand
the observable around the relevant kinematical limits. As in a common
SCET factorisation theorem, this implies that the soft and each
collinear sector is evaluated using the observable expanded in the
corresponding soft or collinear limits.
Performing the explicit expansion of the observables has the advantage
of dropping undesired subleading logarithmic corrections, as well as
all power corrections, which are retained when using the full
observable dependence. This is important in a purely resummed
calculation. Furthermore, in the context of the method presented in
this article, it allows for the numerical extraction of specific
ingredients to other resummation approaches. In the following we will
briefly discuss the implications of the latter choice in the
calculation of the transfer functions defined in the previous section.

\subsection{The soft sector}
The evaluation of the transfer functions involving the soft sector,
namely
eqs.~\eqref{eq:NLL_transfer_final},~\eqref{eq:NNLL_transfer_soft_1},
and~\eqref{eq:NNLL_transfer_soft_2}
can be carried out by taking the soft limit of the observable $V$, namely
\begin{equation}
V(q_n, q_\bn; k_1, \ldots, k_N) \to V_s(q_n, q_\bn; k_1, \ldots, k_N)\,.
\end{equation}

While this choice automatically eliminates any source of
subleading-power corrections in the resummed prediction, for generic
observables it may still lead to the inclusion of subleading
logarithmic effects in the result. As discussed in
Section~\ref{sec:logcounting}, this is a common feature in any
resummation approach. 

Nevertheless, it is interesting to work out a simple recipe to neglect
sources of subleading logarithmic corrections due to the approximation
of the observable.
This can be easily done by noticing that, at NLL, the relevant
kinematics is described by an ensemble of soft radiation ($E_i \ll Q$)
that is strongly separated in angle and collinear to either of the legs
$\ell$ in the hard process. Since $\eta_i \sim \chi_i \ln 1/v$, for
$v\to 0$ the collinear limit implies that all $\chi_i$ are of order 1,
while the strong separation ordering implies that all $\chi_i$'s are
different from each other. In other words, the soft-collinear
approximation for the observable $V_{sc}$ is obtained by expanding
the full observable about the limit
\begin{align}
\label{eq:V_sc_definition}
V_{sc}(q_n, q_\bn; k_1, \ldots, k_N) : \qquad  E_i \ll Q \,, \quad \chi_i \neq \chi_j\,, \quad \chi_i \sim 1
\,.
\end{align}
The corresponding approximation can be adopted in the calculation of
the NLL transfer function
$\fullF^{\rm NLL}$~\eqref{eq:NLL_transfer_final}
and~\eqref{eq:fullF_NNLL_NLL_expanded}, that can be recast as
\begin{align}
\label{eq:obs_expansion_soft}
\fullF^{\rm NLL}\left[P^{\rm NLL}_{s,\ell}, V_s\right] = &\fullF^{\rm
  NLL}\left[P^{\rm NLL}_{s,\ell}, V_{sc}\right]
\nn
& + \left\{\fullF^{\rm
  NLL}\left[P^{\rm NLL}_{s,\ell}, V_s\right] - \fullF^{\rm NLL}\left[P^{\rm NLL}_{s,\ell}, V_{sc}\right]\right\}\,.
\end{align}
The first line provides the correct NLL result, while the second line
contributes at NNLL and beyond. The same approximation can be used in
evaluating the correction $\delta\fullF^{\rm NLL}$ defined in
eq.~\eqref{eq:fullF_NNLL_NLL_expanded}, as well as the single-soft
correction $\delta \fullF^{\rm NNLL}_{s,1}$ at
NNLL~\eqref{eq:NNLL_transfer_soft_1}.

The second line of eq.~\eqref{eq:obs_expansion_soft} can be once again
expanded systematically beyond NLL. At NNLL, the soft sector is
described once again by an ensemble of soft and collinear radiation
strongly separated in angle, and by a single soft emission that can
either be close in angle to any of the above, or be emitted with a
large angle (away from the collinear limit). The two limits can be
promptly translated into two approximations for the observable $V$,
namely:
\begin{itemize}
\item Two of the existing emissions $k_a$ and $k_b$ are soft and collinear, but have similar rapidities
\begin{align}
\label{eq:V_SR_def}
V_{{ sr}}(q_n, q_\bn; k_1, \ldots, k_N; k_a, k_b): 
\qquad  E_i  &\ll Q \,, \quad \chi_i \neq \chi_j\,,\quad \chi_a \sim
               \chi_b \sim 1\,, \quad \chi_i \sim 1\,.
\end{align}
This configuration is irrelevant for event shapes, but it contributes
for observables that explicitly depend on the relative rapidity
between emissions. A typical case is when jet algorithms are involved
in the definition of the observable~\cite{Banfi:2016zlc}.
\item One of the existing emissions $k_a$ is soft and strongly
  separated in angle from all other emissions, but with large angle
\begin{align}
\label{eq:V_WA_def}
V_{wa}(q_n, q_\bn; k_1, \ldots, k_N; a):
\qquad  E_i  &\ll Q \,, \quad \chi_i \neq \chi_j\,, \quad \chi_i \sim
               1\,,\quad \chi_a \sim 0\,.
\end{align}
\end{itemize}
Accordingly, at NNLL, eq.~\eqref{eq:obs_expansion_soft} can be
expanded as 
\begin{align}
\label{eq:obs_expansion_soft_final}
\fullF^{\rm NLL}\left[P^{\rm NLL}_{s,\ell}, V_s\right] &= \fullF^{\rm
  NLL}\left[P^{\rm NLL}_{s,\ell}, V_{sc}\right]
\nn
& + \sum_{K \in \{wa,\,sr\}}\left\{\fullF^{\rm
  NLL}\left[P^{\rm NLL}_{s,\ell}, V_K\right] - \fullF^{\rm
  NLL}\left[P^{\rm NLL}_{s,\ell}, V_{sc}\right]\right\}+\dots
\nn
& = \fullF^{\rm
  NLL}\left[P^{\rm NLL}_{s,\ell}, V_{sc}\right]
\nn
&+ \sum_{K \in \{wa,\,sr\}}\left\{\fullF^{\rm
  NLL}\left[P^{\rm LL}_{s,\ell}, V_K\right] - \fullF^{\rm
  NLL}\left[P^{\rm LL}_{s,\ell}, V_{sc}\right]\right\}+{\cal O}({\rm N^3LL})\,.
\end{align}

Finally, the double-soft NNLL
correction~\eqref{eq:NNLL_transfer_soft_2} involves two correlated
emissions that are close in rapidity, and therefore it should be
naturally evaluated using the approximation $V_{sr}$ of the observable
described above.

We stress that the approximations described in this section are useful
to neglect subleading logarithmic corrections but they are, strictly
speaking, unnecessary. The kinematic expansion described here can be
systematically extended to higher logarithmic orders by progressively
relaxing the strong angular separation and the collinearity constraint for an
increasing number of emissions. The downside of this approach is that,
while systematically extendable to higher orders, it requires working
out the necessary limits for the observable at each new logarithmic
order.

The restrictions on the observable that we just discussed can be
simply understood in SCET terms as follows. In traditional SCET
resummation the NLL result is entirely given in terms of the anomalous
dimensions. Observable dependence in the anomalous dimension only
arises at 1-loop, and therefore only depends on a single
emission. This picture is consistent with taking the strong angular separation
limit. Taking the collinear limit of the observable guarantees that no
finite contributions are generated, which corresponds to setting the
initial conditions to one in a NLL resummation.  Thus, if one is
concerned with generating subleading logarithmic corrections in
addition to the desired NLL terms, the above mentioned limit of the
observable should be taken.

\subsection{The collinear sector and the zero-bin subtraction}
The collinear transfer function at NNLL~\eqref{eq:NNLL_transfer_coll}
involves a set of soft emissions and one collinear
emission. Therefore, the observable must be evaluated in the
corresponding kinematic limit.

As for the soft transfer function, it is possible to make minor
approximations in order to neglect subleading logarithmic corrections
from the final result.
Following the discussion of the previous subsection, in the
computation of the observable the soft emissions can be simply
approximated by their collinear limit and by assuming that they are
strongly separated in angle, hence neglecting corrections of N$^3$LL
order and higher.

The collinear transfer function~\eqref{eq:NNLL_transfer_coll} at NNLL
is then approximated by
\begin{equation}
\delta \fullF^{\rm NNLL}_{n}[V] \simeq \delta \fullF^{\rm NNLL}_{n}[V_{hc}]\,,
\end{equation}
where $V_{hc}$ is defined on a set of soft-collinear emission which
are strongly separated in angle, and a single (hard) collinear emission
$k_a$.

Finally, similar considerations can be made for the corresponding
zero-bin subtraction~\eqref{eq:NNLL_transfer_0bin}, that at NNLL can
be evaluated using the $V_{sc}$ approximation of the observable, i.e.
\begin{equation}
\label{eq:zero_bin_sc}
\delta \fullF^{\rm NNLL}_{n,\, {\rm 0-bin}}[V]\simeq \delta \fullF^{\rm NNLL}_{n,\, {\rm 0-bin}}[V_{sc}] \,,
\end{equation}
where the replacement also implies that one should take the soft limit
of $V_{\rm max}$ in eq.~\eqref{eq:NNLL_transfer_0bin}. The final NNLL
collinear correction is given by the difference of the above two
equations.

\section{Conclusions}
\label{sec:Conclusions}

In this work we outlined a general formalism to resum rIRC safe
observables in soft collinear effective theory. This formalism
combines two quite different approaches to resummation with one
another and results in a very general and powerful tool that allows
one to resum in a relatively straightforward way most observables of
interest to higher logarithmic order. The first approach used is the
standard SCET technique, which commonly relies on the derivation of a
factorization theorem for a given observable, and then uses RG
equations to resum logarithms at all orders in each of the components
of the factorization theorem.
The second approach is the branching formalism, which uses the
factorization properties of squared amplitudes in QCD to model the
soft and collinear radiation at all-orders, often simulated
numerically using a MC algorithm. By combining these two approaches,
our new formalism exploits the strengths of both methods, namely the
generic numerical solution provided by the branching formalism with
the ability to easily and systematically go to higher orders of the
SCET approach.

The main idea behind this method is inspired by numerical resummation
techniques~\cite{Banfi:2004yd,Banfi:2014sua,Banfi:2018mcq} and
consists of deriving the resummation for a simplified version of the
considered observable, which shares the same LL structure with the
full one.
For a suitably chosen simplified observable, this resummation is much
simpler than for the full observable, and can be always handled
analytically. Moreover, the same simplified observable can be adopted
in many cases.
One can then relate the resummation of the full observable to that of
the simplified one multiplicatively via a transfer function, which can
be computed numerically. While this approach was already introduced
in~\cite{Bauer:2018svx} for the thrust distribution, in this paper we
discussed how it can be generalized to any rIRC safe observable. The
main insight that allowed this generalization was the introduction of
a fully differential energy distribution, which allowed us to
formulate the transfer function for any observable.

We described all details of this method, and provided all equations
necessary for an explicit implementation of the calculation of the
transfer function up to NNLL accuracy. 
By collecting the results obtained in the article, we arrive to the
following master formula for the NNLL transfer function for a generic
rIRC safe observable $V$
\begin{align}
\label{eq:master_formula}
\fullF^{\rm NNLL}(\Phi_B; v)  =&~ \fullF^{\rm NLL}\left[P_\ell^{\rm
  NLL}, V_s\right](\Phi_B; v) + \delta \fullF^{\rm
  NNLL}_{s,1}[V_{s}](\Phi_B; v) + \delta \fullF^{\rm
  NNLL}_{s,2}[V_{s}](\Phi_B; v) 
\nn
&+ \sum_{\ell \in \{n,\bn\}}
  \left(\delta \fullF^{\rm NNLL}_{\ell}[V_{hc}](\Phi_B; v) - \delta
  \fullF^{\rm NNLL}_{\ell,\, {\rm 0-bin}}[V_{s}](\Phi_B; v) \right)\,.
\end{align}

As discussed in
Sections~\ref{sec:transfer_fully_differential_explicit_NNLL}
and~\ref{sec:observable}, it can be sometimes convenient to neglect
any source of subleading logarithmic corrections that may originate
from evaluating eq.~\eqref{eq:master_formula} directly. While this
step is, strictly speaking, not necessary, we find it useful to
provide a simple recipe to neglect all sources of N$^3$LL
contamination from eq.~\eqref{eq:master_formula}. To this end, we can
further decompose the above result as follows
\begin{align}
\label{eq:master_formula_2}
\fullF^{\rm NNLL}(\Phi_B; v) = &\fullF^{\rm NLL}\left[P^{\rm LL}_{s,\ell}, V_{sc}\right](\Phi_B; v) +
  \delta\fullF^{\rm NLL}\left[P^{\rm LL}_{s,\ell}, P^{\rm NLL}_{s,\ell},
  V_{sc}\right](\Phi_B; v)
\nn
&+\sum_{K \in \{wa,\,sr\}}\left(\fullF^{\rm
  NLL}\left[P^{\rm LL}_{s,\ell}, V_K\right](\Phi_B; v) - \fullF^{\rm NLL}\left[P^{\rm LL}_{s,\ell}, V_{sc}\right](\Phi_B; v)\right)
\nn
& + \delta \fullF^{\rm
  NNLL}_{s,1}[V_{s}](\Phi_B; v) + \delta \fullF^{\rm
  NNLL}_{s,2}[V_{sr}](\Phi_B; v) 
\nn
&+ \sum_{\ell \in \{n,\bn\}}
  \left(\delta \fullF^{\rm NNLL}_{\ell}[V_{hc}](\Phi_B; v) - \delta
  \fullF^{\rm NNLL}_{\ell,\, {\rm 0-bin}}[V_{sc}](\Phi_B; v) \right) +
  {\cal O}({\rm N^3LL})\,.
\end{align}

While we did not give any explicit results to N$^3$LL and beyond, the
derivation of our results using the language of effective field theory
was presented in a way that should make the systematic extension to
higher logarithmic accuracies obvious.
In future work, we will present applications of this approach to a
range of experimentally relevant observables and provide generic
computer code that can be used by non-experts to obtain the resummed
prediction for a given rIRC safe observable.

\section*{Acknowledgements}
We are grateful to Luca Rottoli for discussions and constructive
comments on the manuscript. The work of PM is supported by the Marie
Sk\l{}odowska Curie Individual Fellowship contract number 702610
Resummation4PS.

\newpage

\appendix

\section{Resummation for $\Sigma_{\rm max}(v)$}
\label{app:Sigma_max_details}

In this appendix we discuss in some more detail the resummation for
the simple observable $V_{\rm max}$ defined in
Section~\ref{sec:simple_observable_def}.

Given the simple form of the measurement
function~\eqref{eq:measurement_function_Vmax}, one is able to find an
analytic result for this observable at all logarithmic orders. From
the definition, it is easy to see that the characteristic scale of the
soft sector is $\mu_s = Q \,v^{1/a}$, while each collinear sector has
$\mu_{J_\ell} = Q\, v^{1/(a+b_\ell)}$.
By observing that the observable definition does not mix soft and
collinear modes, and in particular that there is no kinematic cross
talk between the different modes, we obtain the simple multiplicative
factorisation theorem
\begin{align}
\label{eq:sigma-factorisation-explicit}
\Sigma_{\max}(\Phi_B; v) &=  H(\Phi_B; \mu_H) \notag\\
&\times\int \df v_n\,\df v_\bn\,\df v_s \,{\cal J}_n(\Phi_B; v_n)
                             {\cal J}_\bn(\Phi_B; v_\bn){\cal
                             S}(\Phi_B; v_s)\Theta(\max\{v_n, v_\bn,
  v_s\} < v)\notag\\
& = H(\Phi_B; \mu_H) \left[ \int^v \df v_n {\cal J}_n(\Phi_B; v_n)
                             \right]\,\left[\int^v \df v_\bn {\cal J}_\bn(\Phi_B; v_\bn)
                             \right]\,\left[\int^v \df v_s {\cal S}(\Phi_B; v_s)\right]\,.
\end{align}
The jet and soft functions are defined as
\begin{align}
{\cal S}(\Phi_B; v_s) &= \frac{1}{N_c} {\rm
  Tr}\langle 0| \bar{Y}_{\bn}^{\dagger}Y_{n}^{\dagger}\delta(v_s -V_{\rm max}[\omega_s])Y_{n}  \bar{Y}_{\bn}|0\rangle
  \,,\notag\\
{\cal J}_n(\Phi_B; v_n) &= \frac{(2\pi)^3}{N_c} {\rm Tr} \langle 0|\frac{\slashed{\bar n}}{2}
                          \chi_{n}\delta(v_n -V_{\rm max}[\omega_n]){\bar 
                                        \chi}_{n}|0\rangle
\,,\notag\\
{\cal J}_\bn(\Phi_B; v_\bn) &= \frac{(2\pi)^3}{N_c} {\rm Tr} \langle 0|
                          {\bar \chi_{\bn}}\delta(v_\bn -V_{\rm max}[\omega_\bn]) \frac{\slashed{n}}{2}\chi_{\bn}|0\rangle
\,,
\end{align}
where $Y_n$ denotes a soft Wilson line along the $n$ direction, while
$\chi_n\equiv W_n^\dagger\xi_n$, with $\xi_n$ being a collinear
fermion field after the BPS~\cite{Bauer:2001yt} field redefinition,
and $W_n$ being a collinear Wilson line.
By comparing eq.~\eqref{eq:sigma-factorisation-explicit} and
eq.~\eqref{eq:Sigmamax_fact} we obtain the obvious relations
\begin{align}
H(\Phi_B; \mu_H) &= \abs{C(\Phi_B)}^2,\notag\\
\Sigma_n^{\max}(\Phi_B; v) &= \int \df v_n \,{\cal J}_n(\Phi_B; v_n)
                             \Theta(v_n < v),\notag\\
\Sigma_\bn^{\max}(\Phi_B; v) &= \int \df v_\bn \,{\cal J}_\bn(\Phi_B; v_\bn)
                             \Theta(v_\bn < v),\notag\\
\Sigma_s^{\max}(\Phi_B; v) &= \int \df v_s\, {\cal S}(\Phi_B; v_s)
                             \Theta(v_s < v)\,.
\end{align}

Resummation proceeds as usual by renormalizing each of the above
building blocks up to a given perturbative order. Given the
multiplicative structure of the factorisation theorem, after
renormalisation each of them satisfies an RGE of the type
\begin{equation}
\mu \frac{\df \ln \Sigma^{\max}_F}{\df \mu} = \Gamma_{\rm
  cusp}[\alpha_s(\mu)] \ln \mu + \gamma_{\rm F}[\alpha_s(\mu)]\,,
\end{equation}
with $F = \{s,n,\bn\}$. Its solution yields
\begin{equation}
\ln \frac{\Sigma^{\max}_F(\mu)}{\Sigma^{\max}_F(\mu_F)} =
\int_{\alpha_s(\mu_F)}^{\alpha_s(\mu)} \frac{\df
\alpha}{\beta(\alpha)}\left(\gamma_{\rm F}[\alpha] + \Gamma_{\rm
  cusp}[\alpha]\int_{\alpha_s(\mu_F)}^\alpha \frac{\df
\alpha^\prime}{\beta(\alpha^\prime)}\right)\,,
\end{equation}
where the anomalous dimensions and the boundary conditions
$\Sigma^{\max}_F(\mu_F)$ have the following perturbative expansions
\begin{align}
\label{eq:anomalous_dimensions}
  \Gamma_{\rm cusp}[\alpha_s(\mu)]  &= \frac{\alpha_s(\mu)}{2\pi}\Gamma^{(1)}_{\rm cusp} + \left(\frac{\alpha_s(\mu)}{2\pi}\right)^2\Gamma^{(2)}_{\rm cusp} + \ldots\,,
\nn
 \gamma_F[\alpha_s(\mu)]  &= \frac{\alpha_s(\mu)}{2\pi}\gamma^{(1)}_F+ \left(\frac{\alpha_s(\mu)}{2\pi}\right)^2\gamma^{(2)}_F+ \ldots\,,\nn
\Sigma^{\max}_F(\mu_F) &= 1 + \frac{\alpha_s(\mu_F)}{2\pi}\Sigma^{\max,\,(1)}_F(\mu_F)+ \left(\frac{\alpha_s(\mu_F)}{2\pi}\right)^2\Sigma^{\max,\,(2)}_F(\mu_F)+ \ldots
\end{align}

The logarithmic accuracy is defined in terms of the perturbative order
of the anomalous dimensions and boundary conditions, as summarized in
Table~\ref{tab:LogCountingSCET}. For example, to achieve LL accuracy,
one only needs $\Gamma_{\rm cusp}^{(1)}$, while for NLL accuracy
$\Gamma_{\rm cusp}^{(n)}$ for $n \le 2$ and $\gamma_F^{(1)}$. For
N$^k$LL accuracy, one needs $\Gamma_{\rm cusp}^{(n)}$ for $n \le k+1$,
$\gamma_F^{(n)}$ for $n \le k$ and boundary condition $\Sigma^{\max,\,(n)}_F(\mu_F)$ with
$n \le k-1$.
\begin{table}
\begin{center}
\begin{tabular}{|l|c|c|c|}
\hline
 &   $\Gamma_{\rm cusp}[\alpha_s]$ & $\gamma_F[\alpha_s]$ & $\Sigma^{\max}_F(\mu_F)$ 
 \\\hline\hline
LL & 1 & -- & --
\\ \hline
NLL & 2 & 1 & -- 
\\ \hline
NNLL & 3 & 2 & 1 
\\ \hline\hline
N$^k$LL & k+1 & k & k-1
\\ \hline\end{tabular}
\end{center}
\caption{The table summarizes the order at which the various
  ingredients to the RGE need to be computed for a given resummation accuracy.
  \label{tab:LogCountingSCET}}
\end{table}

An important property of the simple observable defined in
Section~\ref{sec:simple_observable_def} is that the same definition
can be used for the resummation of any rIRC safe observable $V$, with
the only observable dependence being encoded in the $a$ and $b_\ell$
parameters. A consequence of this property is that the anomalous
dimensions and initial conditions~\eqref{eq:anomalous_dimensions}
depend on the observable under consideration only via the same
parameters, which allows for a generic calculation of
$\Sigma_{\rm max}$ at once.

\section{Kinematics and phase space parametrization}
\label{app:Phase_space_parametrization}
In this appendix we report the parametrizations used for the phase
space in each sector of the SCET Lagrangian with the UV regulated by a
cutoff at $Q$, as described in~\cite{Bauer:2018svx}.  The only phase
space we need to compute for the NNLL transfer function is a single
soft or collinear momentum, and a correlated pair of soft momenta.

\paragraph{Single soft emission\\}
The most natural phase space representation in SCET is in terms of the light-cone components and the transverse momentum, giving
\begin{align}
[\df k] & = \frac{\df^4 k}{(2\pi)^4} \, (2\pi) \, \delta(k^2) = \frac{1}{(2\pi)^3} \frac{1}{2} \df k^+ \df k^- \df^2 k_\perp \delta(k^+ k^- - k_\perp^2) 
\nn
& = \frac{1}{8\pi^2} k_\perp \, \df k_\perp \, \df \eta \, \frac{\df \phi}{2\pi} \,\, \Theta(|\eta| <  \ln (Q/k_\perp))
\,,\end{align}
where the 
rapidity is defined in terms of the light-cone components as
\begin{align}
  \eta = \frac{1}{2} \ln \frac{k^-}{k^+}
  \,.\end{align}
In the above equation the rapidity limit comes from
eq.~\eqref{eq:rapidity-regulator-limit} where we have explicitly used
$\Lambda = Q$ as discussed in
Section~\ref{sec:numerical_treatment_UV_regulator}.

The unit 4-vector $n$ defining the light cone decomposition of the momentum $k$
\begin{align}
k_+ = n \tdot k\,, \qquad k_- = \bn \tdot k\,, \qquad n \tdot \bn =2 \,, \qquad n \tdot k_\perp = \bn \tdot k_\perp = 0
\end{align}
is in principle arbitrary. In the following we will separate the phase
space into two regions associated with either of the legs according to
the definition of the simple observable given in Section~\eqref{sec:simple_observable_def}. For
two legs this corresponds to separating into two hemispheres, giving
\begin{align}
[\df k] = \sum_\ell \frac{1}{8\pi^2} k_{\perp, \ell} \, \df k_{\perp, \ell} \, \df \eta_\ell \, \frac{\df \phi_\ell}{2\pi} \,\, \Theta(0 < \eta_\ell <  \ln (Q/k_{\perp, \ell}))
\,.\end{align}

Now one can change the phase space variables from $k_{\perp, \ell}$ to
$v_\ell \equiv \tilde{V}_{s}(k)$, with $\tilde{V}_{s}(k)$ given in
\eq{Vmax_0}. We introduce the rapidity fraction (with support from 0 to 1) as
\begin{align}
\chi_\ell = \frac{\eta_\ell}{\eta_{\max}} \,, \qquad \eta_{\max} = \frac{1}{a+b_\ell} \ln \frac{1}{v_\ell}
\,.\end{align}
This allows us to write
\begin{align}
\label{eq:PhaseSpace1Def}
[\df k] = \sum_\ell \frac{1}{8\pi^2} \frac{1}{a} \frac{\df v_\ell}{v_\ell} \, \df \chi_\ell \, \frac{\df \phi_\ell}{2\pi} \, \eta_{\max} \, k_{\perp, \ell}^2
\,,\end{align}
where $\eta_\ell$ and $\phi_\ell$ are measured with respect to leg $\ell$ and 
\begin{align}
\label{eq:ktsoft_def}
k_{\perp, \ell} \equiv k_{\perp, \ell}[v_\ell, \eta_\ell, \phi_\ell] = Q v_\ell ^{1/a} \exp\left[\frac{b_\ell}{a} \, \chi_\ell \,\eta_{\max}\right]
\,.
\end{align}

\paragraph{Single collinear emission\\}
Starting again from the standard SCET parametrization in terms of the light-cone components and the transverse momentum, we parametrize the collinear momentum as
\begin{align}
[\df k_n] & = \frac{\df^4 k_n}{(2\pi)^4} \, (2\pi) \, \delta(k_n^2) = \frac{1}{(2\pi)^3} \frac{1}{2} \df k_n^+ \df k_n^- \df^2 k_{n,\perp} \delta(k_n^+ k_n^- - k_{n,\perp}^2) 
\nn
& = \frac{1}{8\pi^2} \, k_{n,\perp} \df k_{n,\perp} \, \frac{\df z_n}{z_n} \, \frac{\df \phi_n}{2\pi}
\,,\end{align}
where now unit 4-vector $n$ is defined by the direction of the collinear sector
\begin{align}
  z_n = \frac{\bn \mcdot k_n}{Q}
  \,.\end{align}
It is useful to express the simple observable in the collinear
limit~\eqref{eq:Vmax_coll} as a function of $k_\perp$ and $z_n$, where
$k_\perp$ is defined with respect to the collinear direction $n$. For a
single collinear emission we obtain
\begin{equation}
\label{eq:Vtilde_n}
\vtn(k) =
\left(\frac{k_\perp}{Q}\right)^{a+b_n} z_n ^{-b_n} (1-z_n)^{-b_n}\,.
\end{equation}
By changing the phase space variables from $k_\perp$ to the value of the observable $v_n \equiv \vtn(k)$, one can write
\begin{align}
[\df k_n] = \frac{1}{8\pi^2} \frac{1}{a+b_n} \frac{\df v_n}{v_n} \, \frac{\df z_n}{z_n}  \, \frac{\df \phi_n}{2\pi} \, k_{n,\perp}^2 \, \Theta(0 < z_n < 1)
\,,\end{align}
where
\begin{align}
\label{eq:kperp_coll}
k_{n,\perp} \equiv k_{n,\perp}[v_n, z_n] = Q v_n^{1/(a+b_n)}
  z_n^{b_n/(a+b_n)} (1-z_n)^{b_n/(a+b_n)}
\,.
\end{align}
Note that the limit $z_n < 1$ comes about automatically from the
constraint $\bn\mcdot k_n < Q$ in the collinear sector of
SCET. Conversely, when performing the zero-bin subtraction, the above
limit comes from the extra constraint due to the UV regulator in
eq.~\eqref{eq:rapidity-regulator-limit-coll}, using $\Lambda = Q$ as
discussed there. This implies that the same phase space can be used
both for the collinear contribution and its corresponding zero-bin
subtraction.

\paragraph{2 correlated soft emissions\\}

The phase space of 2 correlated soft emission has 6 independent
variables. They can be parametrized in many different ways. The
simplest parametrization is to use the usual phase space given in
eq.~\eqref{eq:PhaseSpace1Def} for one (say $k_a$) of the two soft
emissions, and parametrize the remaining phase space in terms of the
relative rapidity between the two emissions $\Delta\eta$, the ratio of
the two transverse momenta 
\begin{equation}
\kappa = \frac{k_{\perp, b}}{k_{\perp,a}}\,,
\end{equation}
and the azimuthal angle ($\phi_b$) of the second emission $k_b$. We
find
\begin{align}
\label{eq:corr-PS-0}
[\df k_a][\df k_b] = \sum_{\ell_a}  \sum_{\ell_b}\frac{1}{(8\pi^2)^2} \frac{1}{a}
  \frac{\df v_{\ell_a}}{v_{\ell_a}} \frac{\df
  \kappa}{\kappa} \, \df \Delta\eta \,\df \chi_{\ell_a} \, \frac{\df
  \phi_{\ell_a}}{2\pi} \,\frac{\df \phi_{\ell_b}}{2\pi} \,
  k_{\perp, a}^4 \, \kappa^2 \eta_{\rm max}
\,.
\end{align}

At NNLL accuracy, we are interested in integrating over phase space
regions where the two emissions $k_a$ and $k_b$ are in the same
hemisphere. All configurations in which the two emissions propagate
into opposite hemisphere start contributing at N$^3$LL, and hence we
purposely omit them in our parametrisation.

We therefore approximate \eq{corr-PS-0} by
\begin{align}
\label{eq:corr-PS-1}
[\df k_a][\df k_b] \simeq \sum_{\ell_a  = \ell_b} \frac{1}{(8\pi^2)^2} \frac{1}{a}
  \frac{\df v_{\ell_a}}{v_{\ell_a}} \frac{\df
  \kappa}{\kappa} \, \df \Delta\eta \,\df \chi_{\ell_a} \, \frac{\df
  \phi_{\ell_a}}{2\pi} \,\frac{\df \phi_{\ell_b}}{2\pi} \,
  k_{\perp, a}^4 \, \kappa^2 \eta_{\rm max}
\,.
\end{align}

\section{SCET amplitudes}
\label{app:SCET_squared_amplitudes}

\paragraph{Single soft emission\\}

The tree-level expression of the 1-particle correlated cluster can be obtained directly from the Feynman rules of the soft sector in SCET and is given by
\begin{align}
\left| M_s^{(0)}(k)\right|^2 =   4 \,C_F \, 4 \pi \alpha_s(k_\perp)\, \frac{1}{k_\perp^2}
\,,
\end{align}
where the factor of 4 comes from the fact that there are two diagrams, each of which has a factor $n \tdot \nbar = 2$, and we have used that for on-shell, massless momenta $k^+ k^- = k_\perp^2$. 

The argument of the running coupling constant needs to be of the natural scale of the soft sector, so using \eq{SCET_scaling} of order $Q \, v^{1/a}$ . The only Lorentz and reparametrization invariant scale that can be formed out of a soft momentum is the transverse momentum, which of course has the right scaling. 

\paragraph{Single collinear emission\\}

In SCET, there are two contributions to the emission of a collinear gluon. The first is from the coupling of the collinear gluon to a collinear fermion, as described by the SCET Lagrangian, and the second is from the collinear Wilson line that is required in any SCET operator to preserve gauge invariance. The two contributions are given by the well known SCET expressions

\begin{align}
M_{n,1}(k_n) &= g_s T^A
\frac{\nbar \tdot (p + k_n)}{(p+k_n)^2} 
\left[
n^\mu 
+\frac{\psl^\perp \gamma^\mu_\perp}{\nbar \tdot p} 
+\frac{\gamma^\mu_\perp (\psl^\perp +\ksl_n^{\perp})}{\nbar \tdot( p +k_n)}
- \frac{\psl^\perp(\psl^\perp+\ksl_n^\perp)}
{\nbar \tdot p \, \nbar \tdot (p+k_n)}\nbar^\mu
\right]  
\\
M_{n,2}(k_n) &= g_s  T^A \frac{1}{\bar n\tdot k_n}\nbar^\mu
\,,
\end{align}
where $n$ denotes the light-like direction along which the collinear sector is defined, and the transverse momentum is measured with respect to that direction. Summing these two contributions and taking the square, one obtains
\begin{align}
\left|M_{n,1}(k_n) + M_{n,2}(k_n)\right|^2 = 2 \, C_F \, 4 \pi \alpha_s(k_{n,\perp})  \, \frac{2 + \frac{2 \bn \cdot p}{\bn \cdot k_n} - \frac{\bn \cdot k_n}{\bn \cdot p}}{p \tdot k_n}
\,.
\end{align}
Using $\bn \tdot k_n + \bn \tdot p = Q$ as well as $n \tdot k_n =
k_{n,\perp}^2 /\bn \tdot k_n$ and $n \tdot p = k_{n,\perp}^2 / \bn
\tdot p$, and defining $z_n = \bn \tdot k_n / Q$, this can be written as
\begin{align}
\left|M_{n}^{(0)}(k_n)\right|^2 \equiv \left|M_{n,1}(k_n) + M_{n,2}(k_n)\right|^2 = 2 \, C_F \, 4 \pi \alpha_s(k_{n,\perp})  \, \frac{1+(1-z_n)^2}{k_{n,\perp}^2}
\,.
\end{align}
As described in
Section~\ref{sec:transfer_fully_differential_explicit_NNLL_collinear},
to obtain the final collinear contribution, one has to perform the
zero-bin subtraction on the collinear transfer function. The necessary
squared amplitude is obtained by taking the $z_n \to 0$ limit of the
above equation. This gives
\begin{align}
\left|M_{n,\, {\rm 0-bin}}^{(0)}(k_n)\right|^2 \equiv  2 \, C_F \, 4 \pi \alpha_s(k_{n,\perp})  \, \frac{2}{n \tdot k_n \, \bn \tdot k_n}
= 2 \, C_F \, 4 \pi \alpha_s(k_{n,\perp})  \, \frac{2}{k_{n,\perp}^2}
\,.
\end{align}
 
The argument of the running coupling constant again is the natural
scale of the collinear sector. At NNLL, the transverse momentum
$k_\perp$~\eqref{eq:kperp_coll} in the collinear squared amplitude can
be approximated with the scaling \eq{SCET_scaling}, i.e.
$Q \, v^{1/(a+b_n)}$, with the difference between the two scales being
of N$^3$LL order.

\paragraph{2 correlated soft emissions\\}
The tree level expression for the 2-body amplitude
$M_s(\Phi_B; k_a, k_b)$ in the soft sector of SCET was computed in
\cite{Kelley:2011ng,Monni:2011gb,Hornig:2011iu}. We write
\begin{align}
\left|M_{s,0}(k_a,k_b) \right|^2= \left[4 \pi \alpha_s(k_{\perp,a})\right]\left[4 \pi \alpha_s(k_{\perp,b})\right] \left|\hat M_{s,0}(k_a,k_b)\right|^2
\,,\end{align}
where the three different color structures $C_F^2$, $C_F \, C_A$ and $C_F n_f T_F$ are given by
\begin{align}
\left| \hat M_{s,0}(k_a,k_b) \right|^2_{C_F^2} = & \, 16 \, C_F^2\frac{1}{k_a^+ k_a^- k_b^+ k_b^-} 
\nn
\left| \hat M_{s,0}(k_a,k_b) \right|^2_{C_F C_A} = & \, 4 \, C_F \, C_A \Bigg[2\frac{(k_a^+ k_b^--k_a^- k_b^+)^2}{(k_a^+ + k_b^+)^2(k_a^- + k_b^-)^2(k_a+k_b)^4}
\nn
& \qquad\qquad\qquad + \frac{(k_a^-)^2 k_b^+(2k_a^++k_b^+)+(k_b^-)^2 k_a^+(2k_b^++k_a^+)}{k_a^+ k_a^- k_b^+ k_b^-(k_a^+ + k_b^+)(k_a^- + k_b^-)(k_a+k_b)^2}
\nn
& \qquad\qquad\qquad + 2 \frac{(k_a^+)^2-k_a^+ k_b^++(k_b^+)^2}{k_a^+ k_b^+(k_a^+ + k_b^+)(k_a^- + k_b^-)(k_a+k_b)^2}
\nn
& \qquad\qquad\qquad - \frac{k_a^-(2k_a^++k_b^+)+k_b^-(2k_b^++k_a^+)}{k_a^+ k_a^- k_b^+ k_b^-(k_a^+ + k_b^+)(k_a^- + k_b^-)} \Bigg]
\nn
\left| \hat M_{s,0}(k_a,k_b) \right|^2_{C_F n_f T_F} &= 8 C_F n_f T_F \left[  \frac{(k_a+k_b)^2 (k_a^- + k_b^-)(k_a^+ + k_b^+)- (k_a^+ k_b^- - k_a^- k_b^+)^2}{(k_a+k_b)^4(k_a^- + k_b^-)^2(k_a^+ + k_b^+)^2}\right] 
\,.
\end{align}
When calculating the 2-particle soft correlated cluster the Abelian contribution cancels and one obtains
\begin{align}
\tilde M_{s,0}^2(k_a,k_b) = \left[4 \pi \alpha_s(k_{\perp,a})\right]\left[4 \pi \alpha_s(k_{\perp,b})\right] \hat M_{s,0}^2(k_a,k_b)
\,,
\end{align}
where 
\begin{align}
\label{eq:Mhat_SO}
 \hat M_{s,0}^2(k_a,k_b) =\frac{1}{2}\left| \hat M_{s}^{(0)}(k_a,k_b) \right|^2_{C_F C_A} + \left| \hat M_{s}^{(0)}(k_a,k_b) \right|^2_{C_F n_f T_F}
 \,,
\end{align}
where the factor of $1/2$ takes into account the symmetry factor for having identical gluon fields.

\section{Numerical algorithms}
\label{app:Numerical_algorithms}

In this section we discuss how to generate the transfer function in a
simple numerical algorithm. The algorithms given below are meant as a
guideline for a numerical implementation of this
formalism. Furthermore, as discussed in the main text, several choices
can be made at NNLL, and below we simply limit ourselves to the main
formulae discussed in the text. We stress that more efficient
implementations can be adopted, like the ones proposed in
refs.~\cite{Banfi:2004yd,Banfi:2014sua}, but their discussion is
beyond the scope of this article.

In the algorithms defined in the remainder of this section we adopt
the prescription outlined in Section~\ref{sec:observable}, according
to which the observable in each sector of the SCET Lagrangian is
expanded about its NLL approximation~\eqref{eq:V_sc_definition},
denoted by $V_{sc}$. We stress that this choice does not affect the
generality of the algorithms given below.
\subsection{Results at NLL}
\label{sec:transfer_final_full_observable_NLL}
The starting point of the numerical algorithm is \eq{NLL_transfer_final}. Using the simple expression of the Sudakov factor given in \eq{Delta_LL_3}, one easily finds
\begin{align}
\Delta_s^{\rm LL}(\Phi_B; v, \eps v) &=
                                       \left(\frac{v}{v_1}\right)^{-\sum_\ell
                                       P^{\rm LL}_{s,\ell}(v)}\,
                                       \left(\frac{v_1}{v_2}\right)^{-\sum_\ell
                                       P^{\rm LL}_{s,\ell}(v)} \ldots
                                       \left(\frac{v_n}{\eps
                                       v}\right)^{-\sum_\ell
                                       P^{\rm LL}_{s,\ell}(v)}\notag\\
& = \Delta_s^{\rm LL}(\Phi_B; v, v_1)\Delta_s^{\rm LL}(\Phi_B; v_1,
  v_2)\ldots \Delta_s^{\rm LL}(\Phi_B; v_n, \eps v)   
\,.\end{align}
Furthermore, we use
\begin{align}
\frac{1}{v_i} \sum_\ell P^{\rm LL}_{s,\ell}(v)
  \left(\frac{v_{i-1}}{v_i}\right)^{-\sum_\ell P^{\rm LL}_{s,\ell}(v)} =  \frac{\df}{\df v_i} \left(\frac{v_{i-1}}{v_i}\right)^{-\sum_\ell P^{\rm LL}_{s,\ell}(v)}
  \,.\end{align}
This allows us to write \eq{NLL_transfer_final} as
\begin{align}
& \fullF^{\rm NLL}(\Phi_B; v) =\notag\\
&\qquad\bigg[ \Delta^{\rm LL}_s(v, \eps v)
+ \sum_{N=1}^\infty \prod_{i=1}^N \sum_{\ell} \int_{\eps
  v}^{v_{i-1}} \!\!\!\!\df v_i \, \frac{\df \Delta^{\rm
  LL}_s(v_{i-1},v_i)}{\df v_i} \int \!  \df \chi_i  \,\frac{P^{\rm
  LL}_\ell(v; \chi_i)}{\sum_\ell P^{\rm LL}_{s,\ell}(v)}  \, \int \!
  \frac{\df \phi_i}{2\pi} \notag\\
&\qquad\qquad\qquad \times\Delta^{\rm LL}_s(v_n, \eps v) \Theta(V_{sc}(q_n, q_\bn; k_1, \ldots, k_N) < v)\bigg]
\,,
\end{align}
where we have eliminated the $1/N!$ symmetry factor by ordering the
emissions according to $v_i$, and we have used that $v_0 \equiv v$.

Thus, the emissions with momenta $k_i(v_i, \chi_i, \phi_i)$ have a distribution that is very similar to that of a parton shower algorithm. They involve an evolution variable ($v_i$) that is monotonically decreasing with each emission, together with a splitting function $P^{\rm LL}_{s,\ell}(v; \chi_i)$ and associated Sudakov factor $\Delta^{\rm LL}_s$. The emissions can therefore be generated by the following algorithm:\\
\begin{algorithm}[H]
   Set weight $w = 1$\;
   Start with $i=0$ and $v_0 = v$\;
   \While{{\rm true}}
   {
      Increase $i$ by 1\;
      Generate a random number $r \in [0,1]$\;
      Determine $v_i$ by solving $\Delta_s^{\rm LL}(\Phi_B; v_{i-1}, v_i)= r$\;
      \If{$v_i < \eps v$} {break\;}
      Choose the leg $\ell$ randomly from a flat distribution\;
      Generate $\chi_i \in [0,1]$ and $\phi_i \in [0,2\pi]$ from a flat distribution\;
      Multiply the event weight $w$ by $P^{\rm LL}_{s,\ell}(v, \chi_i) /
      \left[\sum_\ell P^{\rm LL}_{s,\ell}(v)\right]\times n_{\rm legs}$\;
      Determine $k_i = k(v_i, \chi_i, \phi_i)$ and add to the list of emissions\;
   }
   Return the list of momenta $\{k_i\}$ and associate weight $w$\;
\caption{Generating the soft-collinear emissions}
\label{algo:NLLEmissions}
\end{algorithm}


From the emissions generated with Algorithm~\ref{algo:NLLEmissions}
one then immediately computes the NLL transfer function by taking the
average weight of all emissions with the constraint on the observable $V_{sc}(q_n, q_\bn; \left\{k_i \right\}) < v$. This is results in the algorithm\\
\begin{algorithm}[H]
   Set weight $W = 1$, $W_{\rm Sq} = 1$\;
      \For{$i = 1 \ldots N$}
      {
      Generate a set of soft-collinear emissions $\{k_i\}$ with weight $w$ using Algorithm~\ref{algo:NLLEmissions}\;
      \If{$V_{sc}(q_n, q_\bn; \left\{k_i \right\} ) < v$}
      {
      Increase $W$ by $w$\;
      Increase $W_{\rm Sq}$ by $w^2$\;
      }
      }
      Compute $\fullF^{\rm NLL} \pm \Delta\fullF^{\rm NLL}$ from the average value of $W$ and its standard deviation\;
\caption{Computing the NLL transfer function}
\label{algo:NLLTransfer}
\end{algorithm}

\subsection{Results at NNLL}
\label{sec:transfer_final_full_observable_NNLL}
As discussed in Section~\ref{sec:transfer_fully_differential_explicit_NNLL}, there are 3 broad classes of terms contributing to the transfer function at NNLL. The first is $\fullF^{\rm NLL}[P^{\rm NLL}_{s,\ell}, V]$, the second is the contribution from higher order corrections to the soft correlated clusters $\delta \fullF^{\rm NNLL}_{s}[V](\Phi_B; v)$ and the third the collinear transfer function $\delta \fullF^{\rm NNLL}_{n,\bn}[V](\Phi_B; v)$. We will now give the explicit algorithms for these three contributions.

\subsubsection{$\fullF^{\rm NLL}{[{P^{\rm NLL}_{s,\ell}, V}]}(\Phi_B; v)$:
  Higher order terms in the NLL transfer function}
\label{sec:NNLL_corrections_to_FNLL_algorithms}
As discussed in Section~\ref{sec:transfer_fully_differential_explicit_NNLL}, keeping the NLL terms in the expansions made in \eq{splittingExpansion} results in the NLL splitting function given in \eq{NLL_splitting}. While the LL splitting function given in \eq{Pl_exp} only depended on the values of $v$ and $\chi_a$, the NLL splitting function depends on the value of $v$ as well as all three emission variables $v_a$, $\chi_a$ and $\phi_a$. This means that after performing the analytical integration over $\chi_i$ and $\phi_i$, the function still depends on $v_a$. The Sudakov factor is then given by
\begin{align}
&\Delta^{\rm NLL}_s(\Phi_B, v_{i-1}, v_i)
\\
& \qquad = \exp\Bigg[ - \sum_\ell \int_{v_i}^{v_{i-1}} \frac{\df v_a}{v_a} P^{\rm NLL}_{s,\ell}(v, v_a)\Bigg]
\nn 
& \qquad = \exp\Bigg\{ -\sum_\ell \left[ \left( P^{\rm LL}_{s,\ell}(v)  + P_{s,\ell}^{\rm NLL, \beta}(v) \right) \ln \frac{v_{i-1}}{v_{i}}+  \frac{1}{2} \, \frac{\df P^{\rm LL}_{s,\ell}(v)}{\df \ln 1/v} \, \left(\ln^2 \frac{v}{v_{i}}   -  \ln^2 \frac{ v}{v_{i-1}} \right)\right]\Bigg\}
\notag
\,.
\end{align}
In terms of these expressions, the algorithm can be written as\\
\begin{algorithm}[H]
   Set weight $w = 1$\;
   Start with $i=0$ and $v_0 = v$\;
   \While{{\rm true}}
   {
      Increase $i$ by 1\;
      Generate a random number $r \in [0,1]$\;
      Determine $v_i$ by solving $\Delta_s^{\rm NLL}(\Phi_B; v_{i-1}, v_i)= r$\;
      \If{$v_i < \eps v$} {break\;}
      Choose the leg $\ell$ randomly from a flat distribution\;
      Generate $\chi_i \in [0,1]$ and $\phi_i \in [0,2\pi]$ from a flat distribution\;
      Multiply the event weight $w$ by $P^{\rm NLL}_{s,\ell}(v, v_i, \chi_i, \phi_i ) / \left[\sum_\ell P^{\rm NLL}_{s,\ell}(v, v_i)\right]\times n_{\rm legs}$\;
      Determine $k_i = k(v_i, \chi_i, \phi_i)$ and add to the list of emissions\;
   }
   Return the list of momenta $\{k_i\}$ and associate weight $w$\;
\caption{Generating the soft-collinear emissions with NNLL accuracy}
\label{algo:NNLLEmissions}
\end{algorithm}

The solution to the equation $\Delta^{\rm NLL}_s(\Phi_B, v_{i-1}, v_i)
= r$ is now given by
\begin{equation}
  \frac{v_{i}}{v_{i-1}} = \exp\left[ \frac{\sum_\ell P^{\rm NLL}_{s,\ell}(v,
      v_{i-1}) - \sqrt{\left[\sum_\ell P^{\rm NLL}_{s,\ell}(v,
          v_{i-1})\right]^2 - 2 \sum_\ell\frac{ \df P^{\rm LL}_{s,\ell}(v)}{\df \ln 1/v}
        \ln(r)}}{ \sum_{\ell} \frac{\df P^{\rm LL}_{s,\ell}(v)}{\df \ln 1/v} } \right]
  \,.
\end{equation}
The transfer function is then computed according to
Algorithm~\ref{algo:NLLTransfer}, where the emissions are generated as
just described.

\subsubsection{$\delta \fullF_{s}^{\rm NNLL}(\Phi_B; v)$: Higher order corrections to the soft correlated clusters}
The NNLL soft correction derived in
Section~\ref{sec:transfer_fully_differential_explicit_NNLL_soft} is
further divided into two contributions (see
Eq.~\eqref{eq:fullFNNLL_S}). The correction involving a single
emission reads
\begin{align}
& \delta \fullF^{\rm NNLL}_{s,1}[V](\Phi_B; v) = \Delta_s^{\rm LL}(\Phi_B; v)  \left[ 1 + \sum_{N=1}^\infty \frac{1}{N!}\prod_{i=1}^N  \sum_{\ell} \int_{\eps v} \frac{\df v_i}{v_i} \, \int \!  \df \chi_i  \,P^{\rm LL}_{s,\ell}(v; \chi_i) \, \int \! \frac{\df \phi_i}{2\pi} \right]
\nn
& \qquad \times \sum_\ell \int \! \frac{\df v_{a}}{v_a} \int  \!\df \chi_{a} \int  \!\frac{\df \phi_{a}}{2\pi} \, P^{\rm LL}_{s,\ell}(v; \chi_a) \, K \, \frac{\alpha_s^{\rm LL}(\kappa_{a,\perp})}{2\pi}
\nn
& \qquad \times
\Bigg[ \Theta(V_{sc}(q_n, q_\bn; k_1, \ldots, k_N, k_a) < v) -  \Theta(V_{sc}(q_n, q_\bn; k_1, \ldots, k_N) < v) \,  \Theta(\tilde{V}_{s}(k_a) < v)\Bigg]\,
\,,
\end{align}
which can be computed using a simple variant of
Algorithm~\ref{algo:NLLTransfer}:\\
\begin{algorithm}[H]
  Set weight $W = 1$, $W_{\rm Sq} = 1$\;
     \For{$i = 1 \ldots N$}
     {
     Generate a set of soft-collinear emissions $\{k_i\}$ with weight $w$ using Algorithm~\ref{algo:NLLEmissions}\;
     Choose the leg $\ell$ randomly from a flat distribution\; 
     Choose $v_a \in [0, v]$, $\chi_a \in [0,1]$, $\phi_a \in [0,2\pi]$ randomly from a flat distribution\;
     Construct the momentum $k_a$ from the three emission variables\;
     Multiply the event weight $w$ by $P^{\rm LL}_{s,\ell}(v; \chi_a) \, K \, \frac{\alpha_s^{\rm LL}(\kappa_{a,\perp})}{2\pi} \times n_{\rm legs}$\;
     Compute $\Theta \equiv  \Theta(V_{sc}(q_n, q_\bn; \{k_i\}, k_a ) < v) -  \Theta(V_{sc}(q_n, q_\bn; \{k_i\}) < v) \,  \Theta(\tilde{V}_{s}(k_a) < v)$\;
     \If{$\Theta \neq 0 $} 
     {
     Add $w \, \Theta$ to $W$ \;
     Add $w^2$ to $W_{\rm Sq}$  \tcp*[l]{$\Theta = \pm 1$}
     }
     }
     Compute $\delta \fullF^{\rm NLL} \pm \Delta\delta\fullF^{\rm NLL}$ from average value of $W$ and its standard deviation\;
\caption{Computing the NNLL correction $\delta \fullF^{\rm NNLL}_{s,1}[V](\Phi_B; v)  $}
\label{algo:NNLLTransferS1}
\end{algorithm}

The second term, involving two extra emissions, reads
\begin{align}
& \delta \fullF^{\rm NNLL}_{S,2}[V](\Phi_B; v) =  \Delta_s^{\rm LL}(\Phi_B; v)  \left[ 1 + \sum_{N=1}^\infty \frac{1}{N!}\prod_{i=1}^N  \sum_{\ell} \int_{\eps v} \frac{\df v_i}{v_i} \, \int \!  \df \chi_i  \,P^{\rm LL}_{s,\ell}(v; \chi_i) \, \int \! \frac{\df \phi_i}{2\pi} \right]
\nn
& \qquad \times \int \! [\df k_a][\df k_b] \tilde M_{s,0}^2 (k_a, k_b)
\nn
& \qquad \times
\Bigg[ \Theta(V_{sr}(q_n, q_\bn; k_1, \ldots, k_N, k_a, k_b) < v) -  \Theta(V_{sc}(q_n, q_\bn; k_1, \ldots, k_N, k_{ab}) < v) \Bigg]
\,,
\end{align}
where $k_{ab}$ is the massless momentum that is constructed from from
the transverse momentum and rapidity of the vector $k_a+k_b$, and the
squared matrix element is reported in Eq.~\eqref{eq:Mab_def}. As for
the previous one, this correction can be computed with the following
algorithm:\\
\begin{algorithm}[H]
  Set weight $W = 1$, $W_{\rm Sq} = 1$\;
     \For{$i = 1 \ldots N$}
     {
     Generate a set of soft-collinear emissions $\{k_i\}$ with weight $w$ using Algorithm~\ref{algo:NLLEmissions}\;
     Choose the leg $\ell$ randomly from a flat distribution\; 
     Choose $v_{a} \in [0, v]$, $\chi_{a} \in [0,1]$, $\phi_{a}, \phi_b
     \in [0,2\pi]$ randomly from a flat distribution\;
     Choose $\kappa \in [0, 1]$ uniformly and $\Delta\eta \in (-\infty,
     \infty)$;
     Construct the momenta $k_a$, $k_b$ and $k_{ab}$ from the above emission variables\;
     Build the event weight $w$ according to
     eq.~\eqref{eq:weight_correlated}  \;
     Compute $\Theta \equiv  \Theta(V_{sr}(q_n, q_\bn; \{k_i\}, k_a, k_b) < v) -  \Theta(V_{sc}(q_n, q_\bn; \{k_i\},k_{ab}) < v) $\;
     \If{$\Theta \neq 0 $} 
     {
     Add $w \, \Theta$ to $W$ \;
     Add $w^2$ to $W_{\rm Sq}$  \tcp*[l]{$\Theta = \pm 1$}
     }
     }
     Compute $\delta \fullF^{\rm NLL} \pm \Delta\delta\fullF^{\rm NLL}$ from average value of $W$ and its standard deviation\;
\caption{Computing the NNLL correction $\delta \fullF^{\rm NNLL}_{S,2}[V](\Phi_B; v)   $}
\label{algo:NNLLTransferS2}
\end{algorithm}

\subsubsection{$\delta \fullF_{n,\bn}^{\rm NNLL}(\Phi_B; v)$: The
  collinear correction to the transfer function}

The final NNLL correction, derived in
Section~\ref{sec:transfer_fully_differential_explicit_NNLL_soft},
involves the correction arising from the collinear sector of the SCET
Lagrangian. For each collinear sector, one has
\begin{align}
& \delta \fullF^{\rm NNLL}_{n}[V](\Phi_B; v) = \Delta_s^{\rm LL}(\Phi_B; v)  \left[ 1 + \sum_{N=1}^\infty \frac{1}{N!}\prod_{i=1}^N  \sum_{\ell} \int_{\eps v} \frac{\df v_i}{v_i} \, \int \!  \df \chi_i  \,P^{\rm LL}_{s,\ell}(v; \chi_i) \, \int \! \frac{\df \phi_i}{2\pi} \right]
\nn
& \qquad \times \int \! \frac{\df v_a}{v_a} \int  \! \df z_a  \int  \!\frac{\df \phi_a}{2\pi} \, P^{\rm NLL}_{n}(v, z_a) 
\nn
& \qquad \times
\Bigg[ \Theta(V_{hc}(q_n, q_\bn; k_1, \ldots, k_N, k_a) < v) -  \Theta(V_{sc}(q_n, q_\bn; k_1, \ldots, k_N) < v) \,  \Theta(\tilde{V}_{\ell}(k_a) < v)\Bigg]\,
\,.
\end{align}
where $P^{\rm NLL}_{n}(v, z_n)$ is defined in
Eq.~\eqref{eq:hard_collinear_splitting_function}, and
$\tilde{V}_{\ell}$ (with $\ell = \{n,\bn\}$) is defined in
eq.~\eqref{eq:Vtilde_n}.

Once again, the algorithm is a simple adaptation of
Algorithm~\ref{algo:NLLEmissions}:\\
\begin{algorithm}[H]
   Set weight $W = 1$, $W_{\rm Sq} = 1$\;
      \For{$i = 1 \ldots N$}
      {
      Generate a set of soft-collinear emissions $\{k_i\}$ with weight $w$ using Algorithm~\ref{algo:NLLEmissions}\;
      Choose the leg $\ell$ randomly from a flat distribution\; 
      Choose $v_{a} \in [0, v]$, $z_{a} \in [0,1]$, $\phi_{a}
      \in [0,2\pi]$ randomly from a flat distribution\;
      Construct the collinear momentum $k_a$ from the above emission variables\;
      Multiply the event weight $w$ by $\,P^{\rm LL}_{s,\ell}(v; \chi_i)   \times n_{\rm legs}$\;
      Compute $\Theta \equiv  \Theta(V_{hc}(q_n, q_\bn; \{k_i\}, k_a) < v) -  \Theta(V_{sc}(q_n, q_\bn; \{k_i\}) < v) \,  \Theta(\tilde{V}_{\ell}(k_a) < v) $\;
      \If{$\Theta \neq 0 $} 
      {
      Add $w \, \Theta$ to $W$ \;
      Add $w^2$ to $W_{\rm Sq}$  \tcp*[l]{$\Theta = \pm 1$}
      }
      }
      Compute $\delta \fullF^{\rm NLL} \pm \Delta\delta\fullF^{\rm NLL}$ from average value of $W$ and its standard deviation\;
\caption{Computing the NNLL correction $\delta \fullF^{\rm NNLL}_{n}[V](\Phi_B; v)   $}
\label{algo:NNLLTransferC}
\end{algorithm}

The same algorithm can be adopted for the calculation of the zero-bin
subtraction defined in eqs.~\eqref{eq:NNLL_transfer_0bin}
and~\eqref{eq:zero_bin_sc}, provided one replaces the observable
$V_{hc}$ with $V_{sc}$, $\tilde{V}_{\ell}$ with $\tilde{V}_{s}$
defined in eq.~\eqref{eq:Vmax_0}.

\subsection{$\delta\fullF^{\rm NLL}{\left[P^{\rm LL}_{s,\ell}, P_\ell^{\rm
        NLL}, V\right]}(\Phi_B; v)$:
  expansion of $P^{\rm NLL}_{s,\ell}$ in
  $\fullF^{\rm NLL}{[{P^{\rm NLL}_{s,\ell}, V}]}(\Phi_B; v)$}
\label{app:expansion_PNLL}
In Section~\ref{sec:NNLL_corrections_to_FNLL_algorithms}, we worked
out an algorithm for the computation of $\fullF^{\rm NLL}$ in terms of
the NLL $P^{\rm NLL}_{s,\ell}$ ``splitting function'', used for all
emissions that are generated. An alternative possibility, as discussed
in Eq.~\eqref{eq:fullF_NNLL_NLL_expanded}, is to achieve NNLL accuracy
by describing a single emission with the NLL splitting function, while
using LL splitting functions for all other emissions. In the following
we provide a simple algorithm for the evaluation of this correction.

In this case the NLL Sudakov factor can be written as
\begin{align}
&  \Delta^{\rm NLL}_s(\Phi_B, v, \eps v)  =\Delta^{\rm LL}_s(\Phi_B, v, \eps v) \nn
  & \qquad\qquad\times \Bigg\{ 1 - \sum_\ell \int_{\eps v} \frac{\df
    v_a}{v_a} \!\int \!  \df \chi_a \!\int \! \frac{\df
    \phi_a}{2\pi}\delta P^{\rm NLL}_{s,\ell}(v; v_a, \chi_a, \phi_a)
    \Theta(\tilde{V}_{s} (k_a) < v)\Bigg\}
    \,,
\end{align}
where
\begin{align}
\delta P^{\rm NLL}_{s,\ell}(v; v_a, \chi_a, \phi_a) &=  P^{\rm
                                                  NLL}_\ell(v; v_a,
                                                  \chi_a, \phi_a) -
                                                  P^{\rm LL}_{s,\ell}(v;
                                                  \chi_i)\\
&=\left[ \frac{\df P^{\rm LL}_{s,\ell}(v,\chi_{a})}{\df \ln 1/v} \ln \frac{v \, d_\ell \, g_\ell(\phi_{a})}{v_{a}} - \frac{\beta_1}{\beta_0} \, P^{\rm LL}_{s,\ell}(v,\chi_{a}) \frac{\alpha_s^{\rm LL}(\kappa_\perp)}{4\pi}  \ln (1+t) \right] \notag
\,,
\end{align}
and $\kappa_\perp$ and $t$ are defined in~\eqs{Pl_exp}{alphas_def},
respectively. Similarly, one can perform an analogous expansion for
the real contribution, and after some manipulations we obtain
\begin{align}
& 1 + \sum_{N=1}^\infty \frac{1}{N!}\prod_{i=1}^N \sum_{\ell} \int_{\eps v} \frac{\df v_i}{v_i} \!\int \!  \df \chi_i \!\int \! \frac{\df \phi_i}{2\pi}   \,P^{\rm NLL}_{s,\ell}(v; v_i, \chi_i, \phi_i) 
\nn
& \qquad \qquad  = \left[ 1 +  \sum_{\ell} \int_{\eps v} \frac{\df v_a}{v_a} \!\int \!  \df \chi_a \!\int \! \frac{\df \phi_a}{2\pi}\,\delta P^{\rm NLL}_{s,\ell}(v; v_a, \chi_a, \phi_a)\right]
\nn
& \qquad \qquad \qquad \times \left[1 + \sum_{N=1}^\infty \frac{1}{N!}\prod_{i=1}^N  \sum_{\ell} \int_{\eps v} \frac{\df v_i}{v_i} \!\int \!  \df \chi_i \,P^{\rm LL}_{s,\ell}(v; \chi_i)\!\int \! \frac{\df \phi_i}{2\pi}\right]
\,.
\end{align}
Thus, the correction arising from the higher terms in the expansion of
the splitting function~\eqref{eq:fullF_NNLL_NLL} is given by
\begin{align}
& \delta\fullF^{\rm NLL}{\left[P^{\rm LL}_{s,\ell}, P^{\rm NLL}_{s,\ell},
  V\right]}(\Phi_B; v)
\\
& \qquad  = \sum_{\ell} \int_{0} \frac{\df v_a}{v_a} \!\int \!  \df \chi_a \!\int \! \frac{\df \phi_a}{2\pi}\,\delta P^{\rm NLL}_{s,\ell}(v; v_a, \chi_a, \phi_a)
\nn
& \qquad\qquad  \times
\Delta^{\rm LL}_s(\Phi_B, v, \eps v) \left[1 + \sum_{N=1}^\infty
  \frac{1}{N!}\prod_{i=1}^N \sum_{\ell} \int_{\eps v} \frac{\df
  v_i}{v_i} \!\int \!  \df \chi_i \,P^{\rm LL}_{s,\ell}(v; \chi_i)\!\int \! \frac{\df \phi_i}{2\pi}\right]
\nn
&\hspace{-1cm} \qquad\qquad \times
\left[ \Theta(V_{sc}(q_n, q_\bn; k_1, \ldots, k_N, k_a) < v) -
  \Theta(\tilde{V}_{s}(k_a) < v)\Theta(V_{sc}(q_n, q_\bn; k_1, \ldots, k_N)
  < v) \right]\notag
\,.
\end{align}

This can be implemented using the same variant of
Algorithm~\ref{algo:NLLEmissions} used in the computation of $\delta
\fullF^{\rm NNLL}_{s,1}[V](\Phi_B; v)$ and $\delta \fullF^{\rm NNLL}_{n}[V](\Phi_B; v)$:\\
\begin{algorithm}[H]
   Set weight $W = 1$, $W_{\rm Sq} = 1$\;
      \For{$i = 1 \ldots N$}
      {
      Generate a set of soft-collinear emissions $\{k_i\}$ with weight $w$ using Algorithm~\ref{algo:NLLEmissions}\;
      Choose the leg $\ell$ randomly from a flat distribution\; 
      Choose $v_a \in [0, v]$, $\chi_a \in [0,1]$, $\phi_a \in [0,2\pi]$ randomly from a flat distribution\;
      Construct the momentum $k_a$ from the three emission variables\;
      Multiply the event weight $w$ by $\delta P^{\rm NLL}_{s,\ell}(v; v_a, \chi_a, \phi_a) \times n_{\rm legs}$\;
      Compute $\Theta \equiv \Theta(V_{sc}(q_n, q_\bn; \left\{k_i \right\}, k_a) < v) - \Theta(\tilde{V}_{s}(k_a) < v)\Theta(V_{sc}(q_n, q_\bn; \left\{k_i \right\}) < v)$\;
      \If{$\Theta \neq 0 $} 
      {
      Add $w \, \Theta$ to $W$ \;
      Add $w^2$ to $W_{\rm Sq}$  \tcp*[l]{$\Theta = \pm 1$}
      }
      }
      Compute $\delta \fullF^{\rm NLL} \pm \Delta\delta\fullF^{\rm NLL}$ from average value of $W$ and its standard deviation\;
\caption{Computing the NNLL correction $\delta \fullF^{\rm NLL}[P^{\rm
    NLL}_{s,\ell}, V](\Phi_B; v) $}
\label{algo:NNLLTransferExpanded}
\end{algorithm}

\addcontentsline{toc}{section}{References}
\bibliographystyle{JHEP}
\bibliography{paper_formalism}

\end{document}